\documentclass{ieeeaccess}
\usepackage{cite}
\usepackage{amsmath,amssymb,amsfonts,amsthm}
\usepackage{multirow,tabu,booktabs,longtable,array,float}
\usepackage{epstopdf}
\usepackage{algorithmic}
\usepackage{caption}
\usepackage{graphicx}
\usepackage{textcomp}
\usepackage{url}
\usepackage{wrapfig}

\usepackage{color,soulutf8}
\usepackage[algoruled]{algorithm2e}
\usepackage{bm}

\usepackage{makecell, longtable}

\usepackage{stfloats}

\ifCLASSOPTIONcompsoc
    \usepackage[caption=false, font=normalsize, labelfont=sf, textfont=sf]{subfig}
\else
    \usepackage[caption=false, font=footnotesize]{subfig}
\fi

\def\BibTeX{{\rm B\kern-.05em{\sc i\kern-.025em b}\kern-.08em
    T\kern-.1667em\lower.7ex\hbox{E}\kern-.125emX}}
\begin{document}
\history{ACCEPTED VERSION IEEE ACCESS.}
\doi{10.1109/ACCESS.2017.DOI}

\title{Incorporation of frailties into a
non-proportional hazard regression model and its diagnostics for reliability modeling of downhole safety valves}
\author{\uppercase{Francisco Louzada}\authorrefmark{1},
\uppercase{Jos\'e A. Cuminato\authorrefmark{1}, Oscar M.H. Rodriguez\authorrefmark{2}, Vera L.D. Tomazella\authorrefmark{3}, Eder A. Milani\authorrefmark{1}\authorrefmark{4}, Paulo H. Ferreira\authorrefmark{1}\authorrefmark{5}, Pedro L. Ramos\authorrefmark{1},  Gustavo Bochio\authorrefmark{2},  Ivan C. Perissini\authorrefmark{2}, Oilson A. Gonzatto Junior\authorrefmark{1}, Alex L. Mota\authorrefmark{1}, Luis F.A. Alegr\'ia\authorrefmark{2}, Danilo Colombo\authorrefmark{6}, Paulo G. O. Oliveira\authorrefmark{6}, Hugo F.L. Santos\authorrefmark{6}, Marcus V. C. Magalhães\authorrefmark{6}},
}
\address[1]{Institute of Mathematics and Computer Science (ICMC), University of S\~ao Paulo, S\~ao Carlos, SP, Brazil}
\address[2]{S\~ao Carlos School of Engineering (EESC), University of S\~ao Paulo, S\~ao Carlos, SP, Brazil}
\address[3]{Department of Statistics (DEs), Federal University of S\~ao Carlos, S\~ao Carlos, SP, Brazil}
\address[4]{Institute of Mathematics and Statistics (IME), Federal University of Goi\'as, Goi\^ania, GO, Brazil}

\address[5]{Department of Statistics (DEst), Federal University of Bahia, Salvador, BA, Brazil}

\address[6]{Leopoldo Am\'erico Miguez de Mello Research and Development Center (CENPES - Petrobras), Rio de Janeiro, RJ, Brazil}
\tfootnote{
The authors acknowledge the partial financial support of the Agência Nacional do Petróleo, Gás natural e Biocombustível (ANP), Petrobras, and FAPESP.}

\markboth
{Author \headeretal: Preparation of Papers for IEEE TRANSACTIONS and JOURNALS}
{Author \headeretal: Preparation of Papers for IEEE TRANSACTIONS and JOURNALS}

\corresp{Corresponding author: Eder A. Milani (e-mail: edermilani@ufg.br).}

\begin{abstract}
    In this paper, our proposal consists of incorporating frailty into a statistical methodology for modeling time-to-event data, based on non-proportional hazards regression model. Specifically, we use the generalized time-dependent logistic (GTDL) model with a frailty term introduced in the hazard function to control for unobservable heterogeneity among the sampling units. We also add a regression in the parameter that measures the effect of time, since it  can  directly  reflect  the  influence  of  covariates  on  the  effect  of  time-to-failure. The practical relevance of the proposed model is illustrated in a real  problem  based on a data set for  downhole  safety  valves (DHSVs) used in offshore oil and gas production wells. The reliability estimation of DHSVs can be used, among others, to predict the blowout occurrence, assess the workover demand and aid decision-making actions.
\end{abstract}


\titlepgskip=-15pt

\maketitle

\section{Introduction}\label{sec:Introduction}

The idea of modeling time-to-event data is well established in statistics and widely used in the medical sciences (in the context of survival analysis) and engineering (in the context of reliability analysis). In any of these situations, we are interested in representing the distribution of a non-negative random variable $T$, based on one of its representative functions, such as density, cumulative distribution, or the hazard function.

Many authors have chosen to model survival data in the presence of covariates using the hazard function, which is related to its interpretation. The hazard function represents an interesting alternative, since its interpretation is given in terms of the instantaneous failure rate over time. Perhaps the best known model dedicated to hazard modeling is the Cox model \cite{Cox_72},
which  has brought to light this modeling possibility. The Cox's proportional hazards model is quite flexible and used extensively in survival analysis. It can be easily extended to incorporate, for instance, the effect of time-dependent covariates. A strong assumption, and probably the most problematic of this model is that the failure rates of any two individuals are proportional, popularizing the name Cox proportional hazard (PH) model.  The assumption of proportionality of hazards is not always in accordance with the observed reality in the field, which motivates the study and development of models that relax such a hypothesis.

\begin{figure}[h]
    \centering
    \includegraphics[scale=0.5]{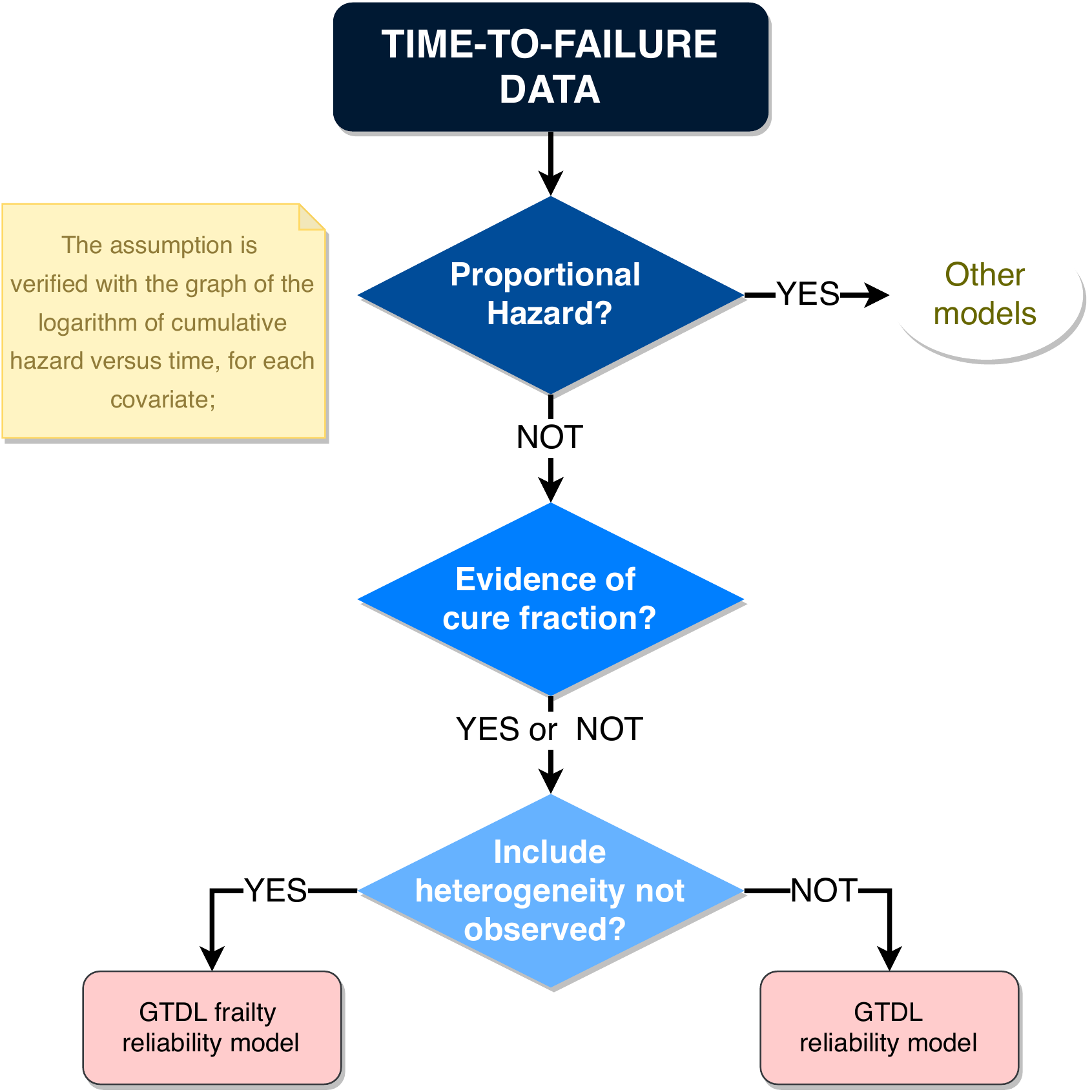}
    \caption{Exemplification of the flexibility of the GTDL and GTDL frailty models.}
    \label{fluxograma}
\end{figure}

Several techniques have been proposed as alternatives to PH modeling. Among others, we can cite 
the use of covariate stratification \cite{kleinbaum2012stratified}, the adoption of time-dependent covariates \cite{kleinbaum2012extension}, 
the nonparametric accelerated failure time model \cite{prentice1978linear}-\cite{kalbfleisch2011statistical},
the hybrid hazard model \cite{etezadi1987extended},
the extension of hybrid hazard models \cite{louzada1997extended}-\cite{louzada1999polyhazard},
and the generalized survival models \cite{liu2017generalized}.
Another approach is the generalized time-dependent logistic (GTDL) model introduced by MacKenzie in \cite{mackenzie1996regression}, whose proposal is to bring a fully parametric competitor for the Cox model. More recently, Louzada-Neto {\it et al.} \cite{louzada2010bayesian} proposed a Bayesian approach to the GTDL model,  Louzada-Neto {\it et al.} \cite{louzada2011interval} compared several techniques for building confidence intervals using parametric and non-parametric resampling methods,  MacKenzie and Peng \cite{mackenzie2014statistical} extended the GTDL model by incorporating a random effect into the hazard function and using h-likelihood procedures that obviate the need to marginalize the risk and survival functions, and Milani {\it et al.} \cite{milani2015generalized} extended the GTDL model by including a gamma frailty term in the modeling. These models have been successfully applied to situations where all units are susceptible to the event of interest, i.e., the presence of a cure fraction in the population is not feasible. Calsavara {\it et al.} \cite{calsavara2020long} proposed an extension of the GTDL model with application in the medical field, where long-term survivors are observed. From the fact that the GTDL and GTDL frailty models have the properties of non-PH and with/without long-term presence (see Figure \ref{fluxograma}), these models can be used in time-to-failure data with these characteristics.

A typical assumption in reliability data analysis is that all of the study units or systems will eventually experience the event of interest if they are followed long enough. Nevertheless, the event may not occur for some units, even after a longer period of time. In manufacturing, for instance, those items that did not fail nor malfunction during the examination time comprise the  cured fraction \cite{vahidpour2016cure}. Nelson \cite{Nelson82} observed the life of insulation on electric motors, which were operated at some levels of temperature; the result is that at low temperature the motors lasted almost indefinitely, while at high temperatures the breake down occurred quickly. From a Bayesian perspective, Lin and Zhu \cite{lin2008cure} proposed a new approach to the reliability analysis of complex systems, where a part of the subsystems is considered ``longevous'' compared  with the entire system. Thus, the system will not fail due to these subsystems.

Hence, usual survival models, such as the Cox PH model or the accelerated failure time model, are not suitable for such cured individuals. As a result, cure models have been developed for manipulation and analysis of survival data with cure fraction. Boag \cite{boag1949} introduced the standard cure rate model, which is the most widely used cure rate model. His objective was to study cases where there was a fraction of cured patients among those who had received treatment for mouth cancer; the modeling of the failure time of the susceptible group was made by adopting the lognormal distribution and assuming the cure probability to be constant. The mixture cure model was further developed in \cite{berkson}
and later studied extensively by various authors \cite{goldman84}, \cite{Farewell_86}, \cite{kuk1992mixture}, \cite{taylor1995semi}, \cite{Maller_Zhou}, \cite{peng_2000}, \cite{banerjee2004parametric}, \cite{Rodrigues}, among others.


Usual models implicitly admit a homogeneous population  for susceptible systems, but explanatory variables can be included to elucidate the observable heterogeneity. Nonetheless, genetic, environmental factors or even information that, for some reason, was not considered in the planning, can cause a portion of the unobserved heterogeneity. Hougaard \cite{hougaard1991modelling} discussed the benefits of adopting two sources of heterogeneity, the observable (given by explanatory variables) and the unobservable, considering for the latter some distribution families. Unobserved heterogeneity can be controlled by introducing a random effect to the hazard function, known as frailty (the term ``frailty'' was introduced in \cite{vaupel1979impact}).
In this situation, the frailty models are widely used; for more details, we refer the reader to \cite{wienke2010frailty}. The exclusion of a relevant explanatory variable in the modeling will increase the amount of unobservable heterogeneity, thus, the frailty makes it possible to evaluate the effects of the explanatory variables that were not considered in the modeling.  Therefore, the frailty, besides explaining the heterogeneity between the systems, also allows to alleviate the absence of important covariates.

The challenges in the construction of oil wells are increasing over time, either due to the increase in technical difficulties due to the greater complexity of the areas to be explored or by the improvements in the rules of regulatory bodies aiming at increasing safety.  The DHSV (downhole safety valve) is a subsurface safety valve, which is positioned in the oil production pipeline column below the seabed; its function is to enable the production column to be closed almost instantly, preventing uncontrolled leakage of hydrocarbons into the environment in the event of a catastrophic wellhead accident. The failure (closing or opening unwantedly and other unexpected actions) of the DHSV generates several unforeseen events causing great financial losses. Demonstrating reliability performance of DHSVs is an important activity related to risk assessment and management of offshore well systems \cite{selvik2015review}.

The study of reliability associated with DHSV contemplates many ramifications, even in statistics itself, including (but not limited to): (i) investigating current failures; (ii) evaluating their root causes, failure mechanisms and effects; (iii) estimating and improving the reliability of its components; (iv) developing degradation models as part of a testing strategy, among others. Selvik and Abrahamsen \cite{selvik2015review} studied and discussed the specific statistics for the period 2002 - 2013, focusing on reliability aspects of the collected data. Their study also included a literature review and some testing data collected directly from oil and gas companies, to provide a more nuanced picture of the reliability issues. Rausand and Vatn \cite{rausand1998reliability} discussed the impact of using a Weibull life distribution instead of an exponential distribution, based on a specific data set for surface-controlled subsurface safety valves used in offshore oil and gas production wells. Oliveira \cite{oliveira2016} compared the reliability of some control systems models taking into account the equipment positions throughout the system and their failure rates with a vision more focused on loss of production than in security. Colombo {\it et al.} \cite{colombo2020regression} analyzed the behavior of several machine learning models to assess the reliability of DHSVs for further comparison against traditional statistical techniques, based on experimental evaluation over a data set collected from a Brazil's oil and gas company. In the context of this study, we would like to identify the association of the failure rate behavior with some environmental variables, such as the hydrogen sulfide (H2S) concentration, temperature, pressure, gas/oil ratio and water column. For this, we use  the GTDL and GTDL frailty models, since the assumption of PH was not validated, consequently the Cox model cannot be used.

After fitting a model, it is necessary to check the validity of its assumptions, as well as to carry out robustness studies to detect possible influential or extreme observations that can provoke distortions in the results of the analysis. There are several works in the survival analysis setting that present such analysis \cite{yiqi2016influence}, \cite{leao2017birnbaum}, \cite{leao2018incorporation}.
In this study, we discuss the global influence starting from case-deletion,
in which the influence of the $i$-th observation on the parameter estimates is investigated by excluding it from the analysis. We propose diagnostic measures based on case-deletion for the GTDL and GTDL frailty regression models, in order to determine which units might be influential in the analysis.
To motivate our research, we describe the following real data set related to DHSVs.

\section{Motivating example in oil and gas industry}

\begin{figure}[!ht]
\centering
\includegraphics[width=0.8\linewidth]{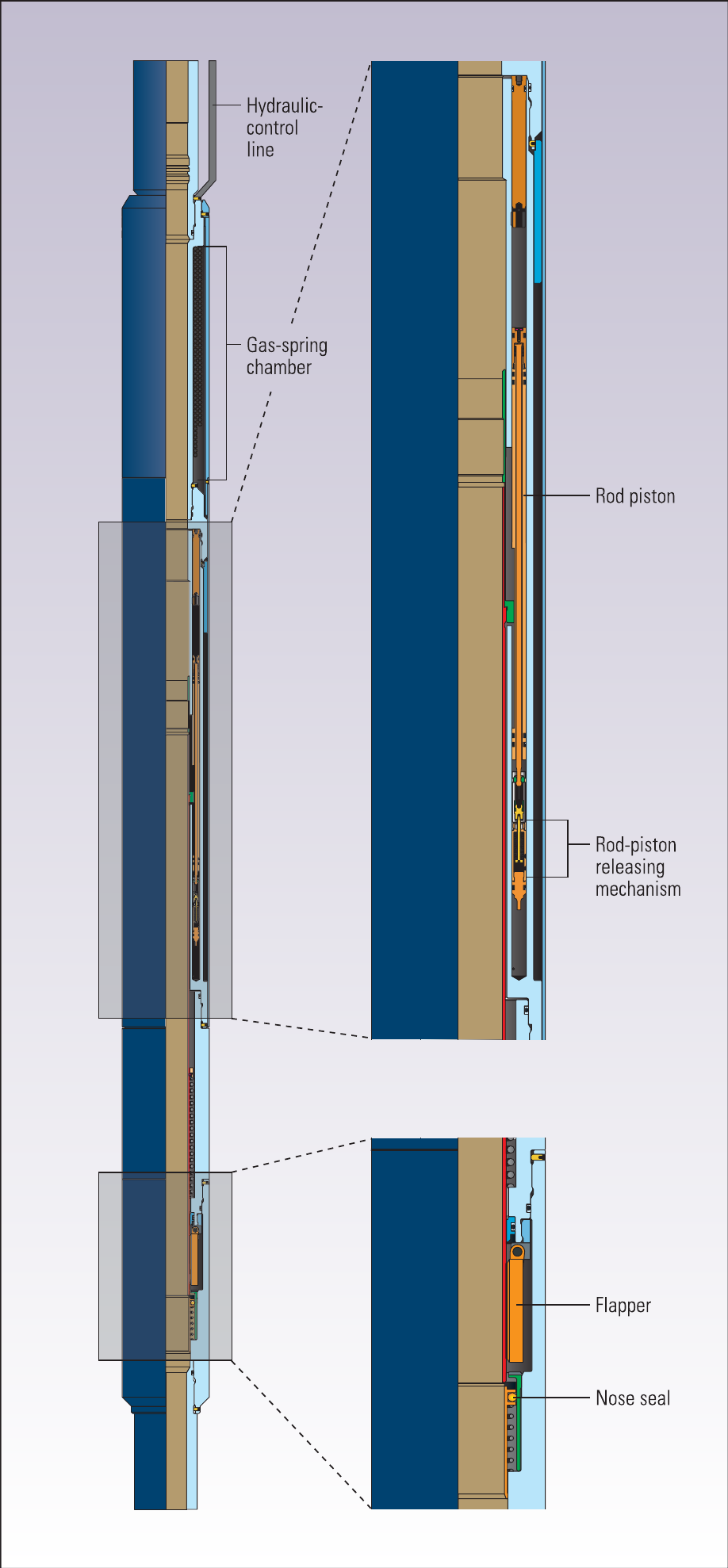}
\caption{Tubing-retrievable charged - downhole  safety  valves (TRC-DHSV) illustration (taken from \cite{garner2002ready}).}
\label{valvula_dhsv}
\end{figure}

The motivation for our study came from a real-world reability  data set corresponding to the DHSVs used in the exploration of Petrobras' (abbreviation of \textit{Petr\'oleo Brasileiro S.A.}) oil wells in Brazil. Illustrated in Figure \ref{valvula_dhsv}, the DHSV is a subsurface safety valve whose function is to prevent uncontrolled leakage of hydrocarbons into the environment in the event of a catastrophic wellhead accident.


The records show the time (in years) of valve's life, whether or not there was a suspension of use and some other explanatory variables, presented in Table \ref{variaveis}, which are divided into groups according to their characteristics. The type of variable is also highlighted. 

\begin{table}[!ht]
\centering
\caption{Explanatory variables divided by group.}
\resizebox{\linewidth}{!}{
\setlength{\tabcolsep}{3pt}
\begin{tabular}{llcc}
  \hline
      Group	&   Variable	           & Abbreviation	&    Type of variable \\
\hline      
Environmental	& Closed well temperature  &	CWT	        &     Continuous \\
                & Closed well pressure	   &    CWP         &     Continuous \\
                & Operating unit	       &     OU         &     Qualitative \\
                &	Water column	       &     WC	        &     Continuous \\
\hline                
Operation	    &  Well flowing pressure   &	WFP	        &     Continuous \\
                &  Flow rate	           &     FR	        &     Continuous \\
\hline                
Valve	        &  Manufacturer            &   	Mfr.       &     Qualitative \\
                &  Family	               &  Family	    &     Qualitative \\
                &  Dimension	           &    Dim. 	    &     Qualitative \\
                & Pressure class	       &     PC	        &     Qualitative \\
\hline                
Flow	        & H2S concentration	       &     H2S	    &     Continuous \\
                & Basic sediment and water &     BSW	    &     Continuous \\
                & Gas oil ratio	           &     GOR	    &     Continuous \\
\hline
\end{tabular}
\label{variaveis} }
\end{table}

\begin{figure}[!ht]
\includegraphics[width=1\linewidth]{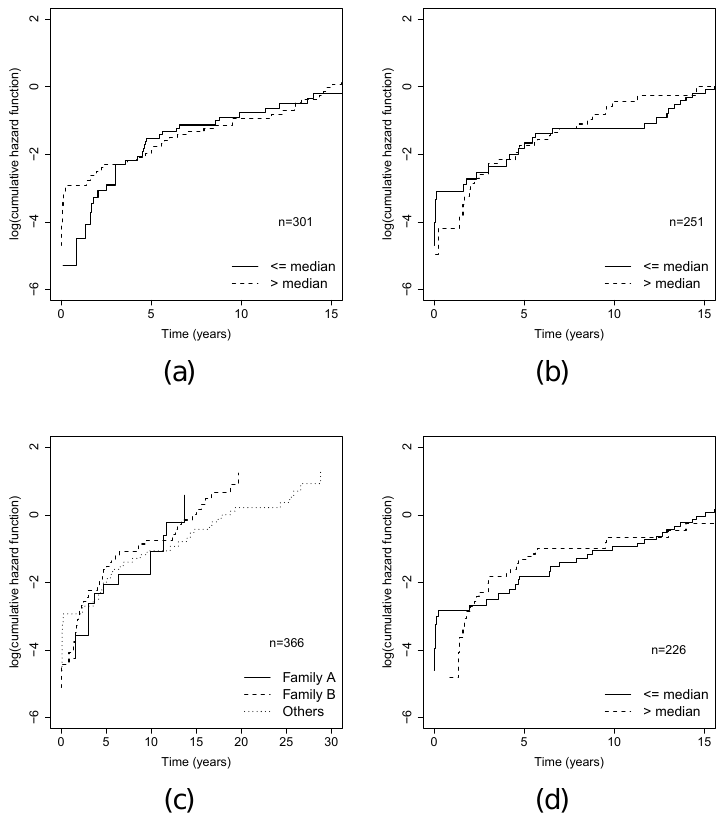}
\caption{Plots of the logarithm of estimated cumulative hazard function \textit{versus} time, for the covariates: (a) CWT, (b) FR, (c) Family, and (d) GOR.}
\label{verificacao_proporcionalidade_1}
\end{figure}

Our objective is to study the time until the failure of the DHSV and to identify possible associations between the characteristics of the valve, the environment, the functioning and the flow with the time-to-failure; for this, we adopt risk modeling. The assumption of PH is verified by means of the graph of the logarithm of cumulative hazard \textit{versus} time, for each covariate; \cite{colosimo2006, collett2015modelling, kleinbaum2012evaluating} present detailed discussions of how to assess the PH assumption. The log-cumulative hazard plots shown in Figure \ref{verificacao_proporcionalidade_1} indicate non-proportionality for the covariates: CWT, FR, Family and GOR. The remaining graphs are presented in Figure \ref{verificacao_proporcionalidade_2} of Appendix.

%

In addition to the graphical verification, we also carried out the investigation of the property of PH through Shoenfeld residuals \cite{grambsch1994proportional}. The null hypothesis is proportionality of hazards. The obtained results are summarized in Table \ref{tab:schoenfeld}. Considering a significance level of 10\%, the variables CWT, FR, Family, Dim. and GOR present non-PH. The global analysis was not performed due to the large amount of missing data in the covariates.

\begin{table}[h]
    \centering
    \caption{Result of the hypothesis test to check the PH assumption.}
    \begin{tabular}{ccccc}
    \midrule
         Variable & p-value &  & Variable & p-value \\
         \midrule
         CWT  &  0.049  & & Family & 0.015 \\
         CWP  &  0.727  & & Dim.   & 0.075 \\
         OU   &  0.513  & & PC     & 0.293 \\ 
         WC   &  0.173  & & H2S    & 0.250 \\
         WFP  &  0.596  & & BSW    & 0.563 \\
         FR   &  0.065  & & GOR    & 0.063 \\
         Mfr. &  0.506  & &        &       \\
         \midrule
    \end{tabular}
    \label{tab:schoenfeld}
\end{table}

Since the results of the analysis of the property of PH are not unanimous, we decided to use the GTDL and GTDL with gamma frailty models. As explained in \cite{mackenzie1996regression}, the GTDL model can approach the PH model. When it occurs, the estimates of the regression parameters  should be similar in both models.


The inclusion of a frailty term in the traditional models can also be needed. As mentioned earlier, unobservable heterogeneity among units or systems may play an important role in assessing reliability, while its omission can cause biased results. Hence, this example serves as a motivation for the joint modeling of heterogeneity among valves by their frailties and the possible presence of a cured fraction of them when predicting their reliabilty.

The remainder of the paper is organized as follows. In Section \ref{background}, we provide further details on the GTDL and GTDL frailty models, such as survival and hazard functions, their cure rate version and inference methods based on the likelihood function. Section \ref{influence} describes and discusses the influence diagnostics based on case-deletion. Section \ref{application} presents the fitted models to the groups of variables and the diagnostic analysis. Finally, Section \ref{conclusions} gives some concluding remarks.

\section{Background}
\label{background}

In this section, we present the GTDL and GTDL with gamma frailty regression models, highlighting the hazard, reliability and probability density functions. These models are useful for data sets with covariates that do not satisfy the proportional hazards assumption.

\subsection{GTDL model}
\label{gtdl_model}
Let $T>0$ be a random variable representing the failure time and $h(t)$ the  instantaneous failure rate or baseline hazard function. According to \cite{mackenzie1996regression}, the hazard function of the GTDL model is given by
\begin{eqnarray}
h_0\left(t \mid {\bm x}\right)=\lambda\displaystyle\frac{\exp\left\{\alpha t+{\bm x}^\top\bm{\beta}\right\}}{1+\exp\left\{\alpha t+{\bm x}^\top\bm{\beta}\right\}},
\label{risco_mackenzie}
\end{eqnarray}
where $\lambda>0$ is a scalar, $\alpha \in \mathbb{R}$ is a measure of the time
effect, ${\bm x}^\top=({ x}_{1},\ldots,{x}_{p})$ are the sets of covariates and $\bm{\beta}$=$(\beta_1,\ldots,\beta_p)^\top$ are the regression coefficients.

The corresponding reliability function, $R(t|{\bm x})$, and probability density function, $f(t|{\bm x})$, are given, respectively, by
\begin{eqnarray}
R\left(t\mid{\bm x}\right)=\left(\frac{1+\exp\left\{\alpha t+{\bm
		x}^\top\bm{\beta}\right\}}{1+\exp\left\{{\bm x}^\top\bm{\beta}\right\}}\right)^{-\lambda/\alpha
}
\label{sobre_mackenzie}
\end{eqnarray}
and
\begin{eqnarray*}
\label{fdp_mack}
f\left(t\mid{\bm x}\right)&=&
\left(\lambda\displaystyle\frac{\exp\left\{\alpha t+{\bm
		x}^\top\bm{\beta}\right\}}{1+\exp\left\{\alpha t+{\bm x}^\top\bm{\beta}\right\}}
\right) \\
&&\times \left(\frac{1+\exp\left\{\alpha t+{\bm x}^\top\bm{\beta}\right\}}{1+\exp\left\{{\bm
		x}^\top\bm{\beta}\right\}}\right)^{-\lambda/\alpha
}.
\end{eqnarray*}

The hazard function (\ref{risco_mackenzie}) has monotonic behavior, being defined by the value of parameter $\alpha$. More specifically, when $\alpha>0$, the hazard function is increasing; when $\alpha<0$, it is decreasing; and finally, when $\alpha=0$, the hazard function is constant over time, that is, the resulting model is a PH model with exponential baseline hazard function, as highlighted in \cite{mackenzie2014statistical}.

The GTDL model is said to be a non-PH model because the ratio of the hazard functions for two individuals is not constant over time. Consider two systems $i$ and $j$, $i\neq j$, with  covariate vectors ${\bm x}_{i}$ and ${\bm x}_{j}$, ${\bm x}_{i} \neq {\bm x}_{j}$, for $i,j=1,\ldots,n$. Then, the ratio of the hazard functions is given by
\begin{eqnarray}
\tau\left(t\mid{\bm x}_{i},{\bm x}_{j}\right)&=&\frac{h_0\left(t\mid{\bm x}_{i}\right)}{h_0\left(t\mid{\bm
		x}_{j}\right)} \nonumber\\
		&=&\frac{1+\exp\left\{\alpha t+{\bm x}_{j}^\top\bm{\beta}\right\}}{1+\exp\left\{\alpha t+{\bm
		x}_{i}^\top\bm{\beta}\right\}} \nonumber \\
		&&\times\exp\left\{\left({\bm x}_{i}-{\bm x}_{j}\right)^\top\bm{\beta}\right\}.
\label{Prova_NP}
\end{eqnarray}
Note from (\ref{Prova_NP}) that the time effect does not disappear, and hence the non-PH condition becomes evident. As mentioned in \cite{mackenzie1996regression}, the GTDL model is neither a PH model nor an accelerated life model.

The reliability function (\ref{sobre_mackenzie}) is proper for $\alpha > 0$, i.e., $R(0| {\bm x})=1$ and $\displaystyle \lim_{t\rightarrow\infty} R(t| {\bm x})=0$. But when the value of parameter $\alpha$ is negative, the GTDL model naturally acquires an improper distribution, i.e., $R(0| {\bm x})=1$ and $\displaystyle \lim_{t\rightarrow\infty} R(t| {\bm x})=p > 0$; hence, the GTDL model is a cure rate model when $\alpha <0$. An advantage of the GTDL model over the mixed model \cite{berkson},
is that the former makes no assumption about the existence of a cure rate, leaving the data to indicate the presence or not of a cure fraction. In literature, models with this property have recently been called ``defective'' \cite{balka2009review}, \cite{rocha2016two} and \cite{scudilio2018defective}. 

In reliability, the event (failure or error) may not occur with some units, even after a very long period of time. Thus, the cure rate in the population is calculated as the limit of the reliability function (\ref{sobre_mackenzie}) when $\alpha<0$, given by
\begin{eqnarray*}
\label{prop_GTDL}
\displaystyle
p({\bm x})=\lim_{t\rightarrow\infty}R\left(t \mid {\bm x}\right)=
%
\left({1+\exp\left\{{\bm x}^\top\bm{\beta}\right\}}\right)^{\lambda/\alpha
}\in(0,1).
\end{eqnarray*}

Let $T_i>0$ be a random variable denoting the failure time for the $i$-th unit, and $\delta_i$ a censoring indicator variable, which is $\delta_i=0$ if the observed time is censored and $\delta_i=1$ otherwise, for $i=1, \ldots, n$. Also, consider $\boldsymbol{\eta}=(\lambda, \alpha, \beta_1, \ldots, \beta_p)$ and assume that $T_i$'s are independent and identically distributed (IID) random variables with hazard and reliability functions specified, respectively, by (\ref{risco_mackenzie}) and (\ref{sobre_mackenzie}). Then, the likelihood function considering right-censored reliability data is given by  
\begin{eqnarray*}
L(\boldsymbol{\eta})&=&\displaystyle\prod_{i=1}^n
h_0\left(t_i\mid{\bm x}_i\right)^{\delta_i}R\left(t_i\mid{\bm x}_i\right)
\nonumber \\
&=&\displaystyle\prod_{i=1}^n \left(\lambda\displaystyle\frac{\exp\left\{\alpha t_i+{\bm x}_i^\top\bm{\beta}\right\}}{1+\exp\left\{\alpha t_i+{\bm x}_i^\top\bm{\beta}\right\}} \right)^{\delta_i} \\
&& \times \left(\frac{1+\exp\left\{\alpha t_i+{\bm x}_i^\top\bm{\beta}\right\}}{1+\exp\left\{{\bm
		x}_i^\top\bm{\beta}\right\}}\right)^{-\lambda/\alpha
}
\end{eqnarray*}
and the log-likelihood function, $\ell(\boldsymbol{\eta})=\log\left( L(\boldsymbol{\eta})\right)$, is given by
\begin{eqnarray*}
\ell(\boldsymbol{\eta})&=& \log(\lambda)\sum_{i=1}^n \delta_i 
+ \sum_{i=1}^n \delta_i\alpha t_i
+\sum_{i=1}^n \delta_i{\bm x}_i^\top\bm{\beta}  \\
&&- \sum_{i=1}^n \delta_i \log\left(1+\exp\left\{\alpha t_i+{\bm x}_i^\top\bm{\beta}\right\}\right) 
\\ &&-\frac{\lambda}{\alpha}\sum_{i=1}^n \log\left(1+\exp\left\{\alpha t_i+{\bm x}_i^\top\bm{\beta}\right\}\right) \\
&&+\frac{\lambda}{\alpha}\sum_{i=1}^n \log\left(1+\exp\left\{{\bm x}_i^\top\bm{\beta}\right\}\right).    
\end{eqnarray*}

\subsection{GTDL frailty model}

The frailty model is characterized by the use of an unobservable random effect, which represents information that cannot or has not been observed, such as environmental factors, or information that has not been considered in planning. In conventional frailty models, the frailty variable is introduced in the modeling of the hazard function, with the aim of controlling the unobservable heterogeneity of the units under study, including the dependence of the units that share the same factors.

Based on the GTDL model, the hazard function of the $i$-th individual with the multiplicative frailty term $v_i$ is given by
\begin{eqnarray*}
h_0\left(t\mid{\bm x}_i,v_i\right)=v_i \displaystyle\frac{\lambda \exp\left\{\alpha 
	t+{\bm x}_i^\top\bm{\beta}\right\}}{1+\exp\left\{\alpha t+{\bm x}_i^\top\bm{\beta}\right\}},
\label{risco_fragilidade}
\end{eqnarray*}
where $v_i$ represents a value of the random variable $V$, and $h_0\left(t|{\bm x}_i,v_i\right)$
is called the conditional hazard function of the $i$-th individual given $v_i$. When $v_i>1$, we have that the individual $i$ is more fragile, and becomes stronger when $v_i<1$; hence, the model's name ``frailty'' (or ``fragility''). It is necessary to adopt a known distribution for the random variable $V$; as it can only assume positive values, the natural candidates are: gamma, inverse Gaussian, Weibull, positive stable and power variance function (PVF) distributions, among others; for more details, see \cite{hougaard1995frailty} and \cite{wienke2010frailty}. In general, the restriction adopted is $\mathbb{E}[V]=1$ and $\mathbb{V}{\rm ar}[V]=\theta$, where $\theta$ is interpretable as a measure of unobserved heterogeneity; this restriction was proposed in \cite{vaupel1979impact}.

In order to make inferences on frailty models, we have some options. For instance,  obtaining the marginal hazard and reliability functions and using the traditional likelihood function; or choosing other methods that obviate the need for marginalization, such as the h-likelihood approach proposed by Ha {\it et al.} \cite{ha2001hierarchical} and used in \cite{ha2010robust}. This paper considers the marginal hazard and reliability functions.

\subsection{The GTDL gamma frailty model}

The GTDL gamma frailty model was proposed in \cite{milani2015generalized}, wherein they added a frailty term to the hazard function in a multiplicative way and assumed a gamma distribution for it, i.e., $V \sim \text{Gamma}\left(1/\theta,1/\theta\right)$.
This parametrization is considered to obtain $\mathbb{E}[V]=1$ and $\mathbb{V}{\rm ar}[V]=\theta$.

The marginal reliability function is given by
\begin{eqnarray}
R\left(t \mid {\bm x}\right)=\left[
1+\frac{\lambda\theta}{\alpha}\log\left(\frac{1+\exp\left\{\alpha t +{\bm
x}^\top \bm{\beta}\right\}}{1+\exp\left\{{\bm x}^\top \bm{\beta}\right\}}\right)
\right]^{-\frac{1}{\theta}}, 
\label{sobre_frag_g}
\end{eqnarray}
the corresponding marginal hazard function is given by
\begin{eqnarray}
h\left(t\mid{\bm x}\right)=\frac{h_0\left({t\mid\bm x}\right)}{\left[1+\frac{\lambda
\theta}{\alpha}\log\left(\frac{1+\exp\left\{\alpha t+{\bm
x}^\top \bm{\beta}\right\}}{1+\exp\left\{{\bm x}^\top \bm{\beta}\right\}}
\right)\right]}
\label{risco_frag_g}
\end{eqnarray}
and, finally, the density function is given by
\begin{eqnarray*}
f\left(t\mid{\bm x}\right)=\frac{h_0\left(t|{\bm x}\right)}
{\left[1+\frac{\lambda
\theta}{\alpha}\log\left(\frac{1+\exp\left\{\alpha t+{\bm
x}^\top \bm{\beta}\right\}}{1+\exp\left\{{\bm x}^\top \bm{\beta}\right\}}
\right)\right]^{(1+1/\theta)}},
\end{eqnarray*}
where $h_0\left(t|{\bm x}\right)$ is the hazard function defined in (\ref{risco_mackenzie}).

The hazard function (\ref{risco_frag_g}) takes unimodal and decreasing forms, therefore, the inclusion of frailty makes the GTDL hazard function more flexible; for more details, see \cite{eder2011}. It is evident that such a hazard function depends on the time, consequently, the GTDL gamma frailty model also accounts for non-PH. When the parameter $\theta \rightarrow 0$, the frailty model approaches the traditional GTDL model. As advocated by Aalen {\it et al.} \cite{aalen2015understanding}, heterogeneity is ubiquitous and if ignored, it can lead to misleading comparisons of hazard rates. Therefore, it is prudent to always carry out checks to detect the presence of unobserved heterogeneity. On the other hand, the reliability function (\ref{sobre_frag_g}) behaves similarly to the GTDL model, with cure fraction, $p(\bm{x})$, given by
\begin{eqnarray}
p({\bm x})=\left[1-\frac{\lambda\theta}{\alpha}\log\left(1+\exp\left\{{\bm
x}^\top\bm{\beta}\right\}\right)\right]^{-\frac{1}{\theta}}\in(0,1). \nonumber
\end{eqnarray}

The explanatory variables can be incorporated in the model through the hazard function (\ref{risco_frag_g}) and the scale parameter $\alpha$. The use of regression in the $\alpha$ parameter is a more flexible approach, since it can directly reflect the influence of covariates on the effect of time-to-failure. Due to fhe fact that the parameter $\alpha$ can be estimated to be negative or positive, the identity link function is used, i.e., 
 \begin{eqnarray*}
 	\alpha\left(\mathbf{x_*}\right)=\mathbf{x_*}^{\top}\bm{\alpha},
 \end{eqnarray*}
where $\mathbf{x_*}^{\top}=(1, x_{*_{1}}, x_{*_{2}}, \ldots, x_{*_{q}})$ are the sets of covariates and $\bm{\alpha}=(\alpha_0,\alpha_1, \ldots, \alpha_q)^{\top}$ and are the regression coefficients. In practice, the covariate vectors may be the same, i.e., $\mathbf{x}=\mathbf{x}_*$. Note that we can include the intercept in the vector ${\bm{x}}$, and with that, we also include the parameter $\beta_0$.
MacKenzie \cite{mackenzie2002logistic} presents a discussion of the parameters $\beta_0$ and $\lambda$; in short, only one parameter is allowed for the model to be identifiable, that is, the parameters $\beta_0$ and $\lambda$ are interchangeable.
We chose to include the parameter $\beta_0$ because we are interested in the interpretation of the explanatory variables. Hence, the model parameters are $\boldsymbol{\nu}=(\alpha_0, \ldots, \alpha_q, \beta_0, \ldots, \beta_p, \theta)$. 

Let $T_i>0$ and $\delta_i$ be as previously defined, for $i=1,\ldots,n$. Also, consider that $T_i$'s are IID random variables with reliability and hazard functions given, respectively, by (\ref{sobre_frag_g}) and (\ref{risco_frag_g}). Then, the likelihood function considering right-censored reliability data is given by
\begin{eqnarray*}
L(\boldsymbol{\nu})&=&\displaystyle\prod_{i=1}^n
h\left(t_i\mid{\bm x}_i, \mathbf{x_{*_{i}}}\right)^{\delta_i}R\left(t_i\mid{\bm x}_i, \mathbf{x_{*_{i}}}\right)
\nonumber \\
&=&\displaystyle\prod_{i=1}^n
h_0^*(t_i)^{\delta_i} \left[1+\frac{\theta}{\mathbf{x_{*_{i}}}^{\top}\bm{\alpha}}\log\left(h_0^*(t_i)\right)
\right]^{-\left(1/\theta+\delta_i\right)}.
\label{vero_gtdl_fra}
\end{eqnarray*}
where $h_0^*(t_i)=\frac{\exp\left\{\mathbf{x_{*_{i}}}^{\top}\bm{\alpha} t_i+{\bm
x}_i^{\top}\bm{\beta})\right\}}{1+\exp\left\{\mathbf{x_{*_{i}}}^{\top}\bm{\alpha} t_i+{\bm x}_i^{\top} \bm{\beta}\right\}}$. The log-likelihood function, $\ell({\bm \nu}) =\log\left(L( {\bm \nu})\right)$, is given by
\begin{eqnarray*}
\ell(\boldsymbol{\nu})&=&\sum_{i=1}^n \delta_i\mathbf{x_{*_{i}}}^{\top}\bm{\alpha} t_i +\sum_{i=1}^n\delta_i {\bm x}_i^{\top}\bm{\beta} \\
&& -\sum_{i=1}^n\delta_i \log\left(1+\exp\left\{\mathbf{x_{*_{i}}}^{\top}\bm{\alpha} t_i +{\bm x}_i^{\top}\bm{\beta}\right\}\right)
\nonumber \\
&&-\sum_{i=1}^n\left(\delta_i+\frac{1}{\theta}\right)\log\left(h_0^*(t_i)\right). \label{log_vero2}
\end{eqnarray*}

The maximum likelihood estimates (MLEs) for the parameters of the GTDL and GTDL gamma frailty models are obtained by numerical maximization of the corresponding log-likelihood functions. In this work, we use the \texttt{optimr} function of the ``optimx'' package \cite{optimx1}, \cite{optimx2}, which is implemented in software R Core Team \cite{R1}.

\section{Diagnostic analysis} \label{influence}

In this section, we present two important tools to check the quality of the model's fit to the data. The residual analysis is performed using the randomized quantile residuals. The other tool is the global influence analysis.

\subsection{Randomized quantile residuals}

The randomized quantile (RQ) residuals were proposed in \cite{dunn1996randomized}. They are used to check the overall fit quality of the model. As highlighted in \cite{rigby2005generalized}, the RQ residuals can be used in the presence of censored data. They are also widely used in generalized additive models for location, scale and shape (GAMLSS). The RQ residuals are defined by
\begin{eqnarray}
r_{i}=\Phi^{-1}\left(\widehat{R}(t_i \mid{\bm x}_i)\right), \quad i=1,2,\ldots,n,
\end{eqnarray}
where $\Phi^{-1}(.)$ is the inverse of the cumulative distribution function (or quantile function) of the standard normal and $\widehat{R}(t_i \mid{\bm x}_i)$ is the estimate of the reliability function obtained using the MLEs for the parameters. In this work, we analyze the RQ residuals using the quantile-quantile plot (QQ-plot).

\subsection{Global influence}
Global influence analysis consists of studying the effect of case-deletion from the data. It was introduced by Cook \cite{cook1977detection} and studied later by several authors \cite{cook1982residuals}, \cite{cook1986assessment},  \cite{yiqi2016influence}, \cite{leao2018incorporation}, among others. 
We denote by the subscript ``$(i)$'' the removal of the $i$-th observation from the original data set. The log-likelihood function of the parameter vector $\boldsymbol{\nu}$ is denoted by $\ell(\boldsymbol{\nu})$, as previously given; when we delete the $i$-th observation, we represent it by $\ell_{(i)}(\boldsymbol{\nu})$, with the respective MLE given by $\hat{\boldsymbol{\nu}}_{(i)}=(\hat{\boldsymbol{\alpha}}_{(i)}, \hat{\boldsymbol{ \beta}}_{(i)}, \hat{\theta}_{(i)})^\top$. In this study, we analyze two measures of global influence. The first is the generalized Cook's distance (GD), whose idea is to compare $\hat{\boldsymbol{\nu}}$ and $\hat{\boldsymbol{\nu}}_{(i)}$; if the deleted observation seriously influences the estimates, more attention should be paid to that observation. The GD is given by
\begin{eqnarray*}
\text{GD}_i\left(\boldsymbol{\nu}\right)=\left(\hat{{\boldsymbol \nu}}_{(i)}-\hat{{\boldsymbol \nu}}\right)^\top \left[\Sigma\left(\hat{\boldsymbol \nu}\right)\right]^{-1}\left(\hat{{\boldsymbol \nu}}_{(i)}-\hat{{\boldsymbol \nu}}\right),
\end{eqnarray*}
where ${\Sigma\left(\hat{\boldsymbol \nu}\right)}$ is the expected Fisher information matrix. For the models under study, this matrix is extremely complex, so in practice we use its observed version. An alternative way is to assess $\text{GD}_i\left({\boldsymbol \alpha}\right)$, $\text{GD}_i\left({\boldsymbol \beta}\right)$ and $\text{GD}_i\left(\theta\right)$, whose values reveal the impact of the case-deletion on the estimates of  ${\boldsymbol \alpha}$, ${\boldsymbol \beta}$ and $\theta$, respectively.

The second measure adopted here is the likelihood distance (LD), whose idea is to compare $\ell\left(\hat{{\boldsymbol \nu}}\right)$ and $\ell\left(\hat{{\boldsymbol \nu}}_{(i)}\right)$, similarly to the previous one; if the deleted observation seriously influences the value of the log-likelihood function, it deserves further attention. The LD is given by
\begin{eqnarray*}
\text{LD}_i\left(\boldsymbol{\nu}\right)= 2\left[\ell\left(\hat{{\boldsymbol \nu}}\right)-\ell\left(\hat{{\boldsymbol \nu}}_{(i)}\right)\right].
\end{eqnarray*}

In order to investigate the impact of the detected influential cases, we calculate the relative change (RC), which is computed from parameter estimates with and without removing the influential cases, as follows:
\begin{eqnarray*}
\mbox{RC}_{\nu_{j(i)}}&=&\left | \frac{\hat{\nu}_j-\hat{\nu}_{j(i)}}{\hat{\nu}_j}\right |\times 100\% ,\\
\mbox{RC}_{\mbox{SE}(\nu_{j(i)})}&=&\left | \frac{\mbox{SE}(\hat{\nu}_j)-\mbox{SE}(\hat{\nu}_{j(i)})}{\mbox{SE}(\hat{\nu}_j)}\right |\times 100\%, \\
\end{eqnarray*}
where $\hat{\nu}_{j(i)}$ and $\mbox{SE}(\hat{\nu}_{j(i)})$ are the MLEs and their respective estimated standard errors (SEs) when the $i$-th case is deleted, with $j=1, \ldots, p+q+3$ and $\nu_1=\alpha_0, \ldots, \nu_{q+1}=\alpha_q$, $\nu_{q+2}=\beta_0, \ldots, \nu_{p+q+2}=\beta_p$ and $\nu_{p+q+3}=\theta$. Note that this section was developed for the GTDL gamma frailty model, but if the adopted model is the GTDL one, it is only necessary to remove the parameter $\theta$.

\section{Application} \label{application}

In this section, we apply the proposed methodology to the DHSV data set. The data set is confidential due to the interests of the Petrobras company. The analyses were performed with the R software \cite{RCoreTeam}. A descriptive summary of the failure times or censoring (in years) provides the following main sample results: $n = 366$, ${\rm mean} = 5.0761$, ${\rm median} = 3.6082$, ${\rm minimum} = 0.0164$ and ${\rm maximum} = 28.8000$, with only $83$ $(22.68\%)$ failure times, while the rest are censored times.

For better understanding, we present a descriptive statistical summary of the explanatory variables in Tables \ref{continuas} and \ref{qualitativas}. Table \ref{continuas} displays the minimum value (Min), median, mean, standard deviation (SD), coefficient of variation (CV), skewness (Sk) and kurtosis (K), maximum value (Max) and number of observations ($n$), for the continuous variables. The summary of failure times that are analyzed together with the covariates is also presented in this table. For the qualitative (categorical) variables, it is possible to observe the categories and the number of observations per category in an absolute and relative way, as shown in Table \ref{qualitativas}.

\begin{table}[!h]
\centering
\caption{Descriptive summary of the continuous explanatory variables.}
\resizebox{\linewidth}{!}{
\setlength{\tabcolsep}{3pt}
\begin{tabular}{llccccccccc}
  \hline
Characteristic & Variable & Min & Median & Mean  & SD     & CV   & Sk    & K    & Max   &   $n$ \\
\hline
Flow          & Failure time     & 0.87 & 3.03    & 4.04   & 2.67   & 0.66 & 0.27  & 0.28 & 11.32 & 21 \\
              & Censoring time  & 0.24 & 2.55    &  2.82  &  1.50  & 0.53 &  0.17 & 0.28 & 6.97  & 77 \\
               &  H2S     & 0.00 & 2.45    & 18.57  & 34.14  & 1.84 & 0.38  & 0.03 & 90.00  & 98 \\
               &  BSW     & 0.00 & 1.95    & 26.89  & 37.65  & 1.40 & 0.94  & 0.30 & 100.00 & 98 \\
               &  GOR     & 0.00 & 234.70  & 203.54 & 123.03 & 0.60 & -0.62 & 0.28 & 612.30 & 98 \\
\hline
Environment      & Failure time     & 0.02 & 5.72 & 8.21 & 6.87 & 0.84 & 0.40 & 0.33 & 26.60 & 65 \\
                 & Censoring time  & 0.02 & 4.01 & 4.81 & 3.32 & 0.69 & 0.20 & 0.24 & 22.42 & 154 \\
               & CWT     & 0.00 & 10.00   & 20.00  & 22.89  & 1.14 & 0.62 & 0.33 & 109.00 &219\\
               & CWP     & 0.00 & 0.58    & 0.81   & 0.81   & 0.99 & 0.27 & 0.18   & 3.92 & 219 \\
               & WC      & 0.09 & 0.98    & 1.16   & 0.75   & 0.65 & 0.40 & 0.42   & 2.25 & 219 \\
\hline
Operation      & Failure time     &  0.03 & 3.02 & 3.79 & 2.89 & 0.76 & 0.13 & 0.21 & 11.32 & 29 \\
               & Censoring time  &  0.88 & 3.13 & 3.78 & 2.31 & 0.61 & 0.24 & 0.26 & 13.38 & 151 \\
               & WFP     & 0.73 & 4.21    & 3.75   & 1.33   & 0.35 & -0.55& 0.23    & 6.71 & 180 \\
               & FR    & 0.00 & 2.83    & 2.93   & 2.40   & 0.82 & -0.15& 0.31    & 10.50 & 180 \\
\hline
Valve      & Failure time      &0.03 & 3.00 & 3.50 & 2.84 & 0.81 & 0.10 & 0.21 & 11.32 & 34 \\
           & Censoring time   &0.10 & 3.20 & 3.94 & 2.84 & 0.72 & 0.21 & 0.23 & 14.05 & 258 \\
\hline
\end{tabular} }
\label{continuas}
\end{table}

\begin{table}[!h]
\centering
\caption{Descriptive summary of the qualitative explanatory variables.}
\resizebox{\linewidth}{!}{
\setlength{\tabcolsep}{3pt}
\begin{tabular}{llccc}
  \hline
Characteristic & Variable & \multicolumn{3}{c}{Group/Basin} \\
\hline
Environment      & OU        & Campos (CB) &  Santos (SB) & Espírito Santo (ES)            \\
               &           & 116 (52.96\%)    &  91 (41.55\%)     & 12  (5.49\%)     \\
\hline
Valve          & Mfr.      &  A              &   B               & Others    \\
               &           &  88 (30.14\%)   &   174 (59.59\%)    & 30 (10.27\%)     \\
               & Family   &  A                &  B                & Others    \\
               &           &   87 (29.79\%)   &  150 (51.37\%)    & 55  (18.84\%)     \\
               & Dim.  &   4.5''         &  5.5''           &    \\
               &           & 111 (38.01\%)   &  181 (61.99\%)    &      \\
               & PC &  5,000          &  7,500           & 10,000   \\
               &         & 61 (20.89\%)    &  49 (16.78\%)     & 182 (62.33\%)     \\
\hline
\end{tabular} }
\label{qualitativas}
\end{table}

The database contains 366 observations, but there is a lot of missing data in the explanatory variables. The removal of the cases with missing data reduces the database to only 54 observations, which makes multivariate analyses difficult. Hence, we decided to fit a model for each group of explanatory variables, thereby eliminating the observations with missing data inside the groups. The summary measures previously presented already consider this deletion.

In order to choose/select the explanatory variables, we initially adopted the GTDL model with gamma frailty term and the following steps:

\begin{description}
\item[Step 1: ] \ For each group, select the significant covariates in the ${\bm x}^\top \bm{\beta}$ structure, using the stepwise method and the generalized likelihood ratio test, and also considering $\alpha$ as a scalar;
\item[Step 2: ] \ For each group, select the significant covariates in the $\mathbf{x_*}^{\top}\bm{\alpha}$ structure, using the stepwise method and the generalized likelihood ratio test, considering for the ${\bm x}^\top\bm{\beta}$ structure the covariates obtained in Step 1.
\end{description}

At the end of Step 2, we perform a hypothesis test to verify whether there is observed heterogeneity ($H_0: \theta=0$). In this case, the generalized likelihood ratio test was adopted with the modification presented in \cite{Maller_Zhou}, because the value of the parameter under the null hypothesis is on the boundary of the parametric space. If the null hypothesis is not rejected, the adopted model will be the GTDL; otherwise, the GTDL model with gamma frailty term will be the chosen one.

\subsection{Adjustment to the flow characteristics group}

After applying the procedure previously described, we obtain in Step 1 that the statistically significant explanatory variables are H2S and BSW, while in Step 2 we do not identify any significant variables, and so only $\alpha_0$ is included in the model.  We performed the hypothesis test for the parameter that measures the unobserved heterogeneity 
and obtained a p-value greater than the significance level of 10\%, so the null hypothesis is not rejected and the GTDL model (without frailty) is the adopted one.

The MLEs, SEs and 90\% confidence intervals (90\% CIs) for the GTDL model parameters are presented in Table \ref{emv_fluxo}. By analyzing the confidence intervals, we conclude that all parameters are significant at the 10\% level.

\begin{table}[!h]
\centering
\caption{Estimation results of the GTDL model fitted to the flow characteristics group.}
\resizebox{\linewidth}{!}{
\setlength{\tabcolsep}{3pt}
\begin{tabular}{l|ccc}
  \hline
   Parameter               & MLE     & SE       & 90\% CI \\
  \hline
$\alpha_0$        & 0.7709  & 0.1969   & (0.4470;   1.0948)\\
$\beta_0$         & -5.5598 & 0.8784   & (-7.0048; -4.1148)\\
$\beta_1$ (H2S)   & 0.0362  & 0.0084   & (0.0224;   0.0500) \\
$\beta_2$ (BSW)   & -0.0202 & 0.0121   & (-0.0401; -0.0003)\\
   \hline
\end{tabular}  }
\label{emv_fluxo}
\end{table}

\begin{figure}[h]
\centering
\includegraphics[width=0.5\linewidth]{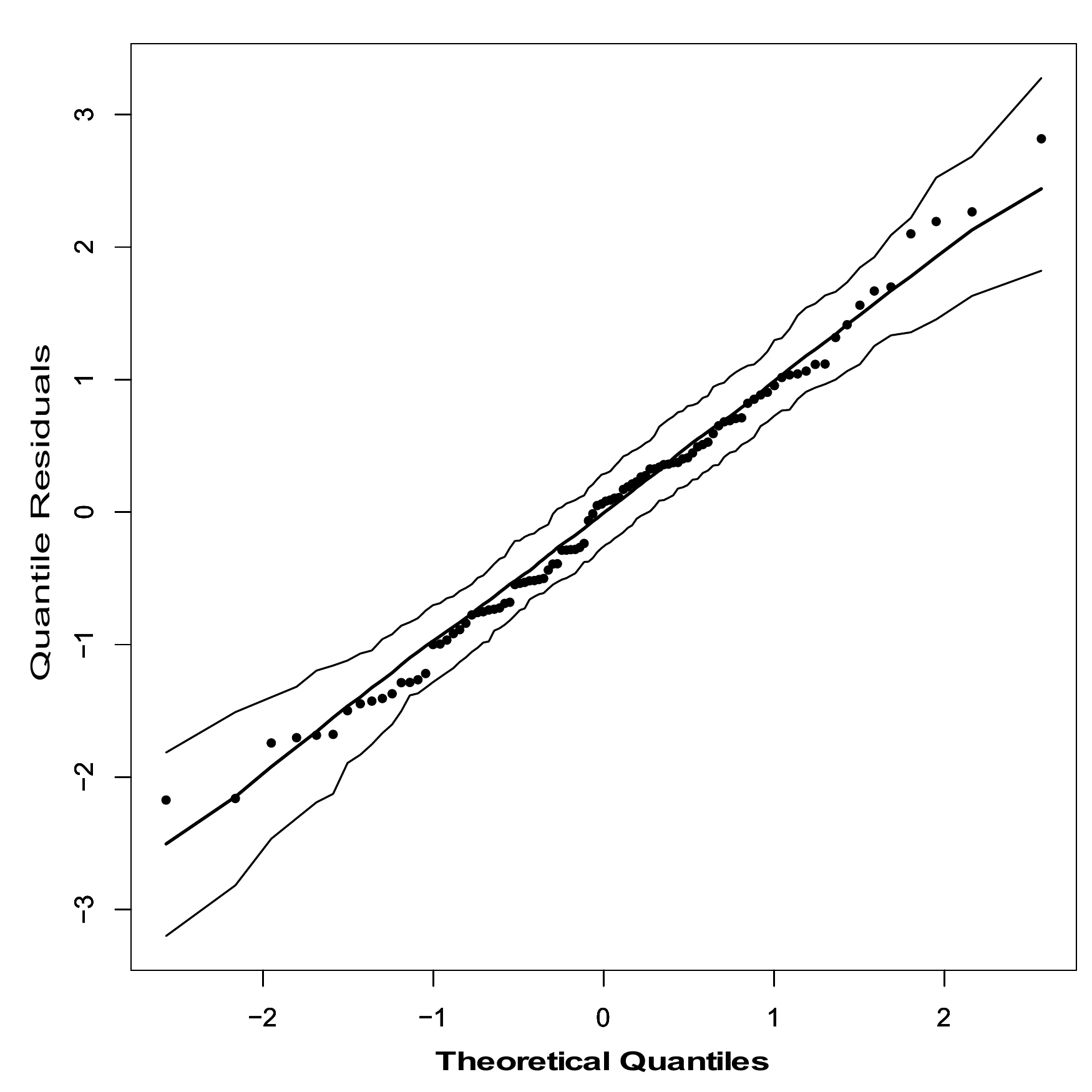}
\caption{QQ-plot with envelope of 95\% for the RQ residuals of the adjustment to the flow characteristics group.}
\label{residuo_fluxo}
\end{figure}

The QQ-plot of the RQ residuals is shown in Figure \ref{residuo_fluxo}. We observed a linear behavior of the residuals (with intercept 0 and slope 1), thus indicating an agreement between the residuals and the standard normal distribution.

For the reliability functions and hazard ratios (HRs) that will be illustrated hereinafter in the text, we adopt the median value of continuous explanatory variables whenever necessary. When the objective is to present reliability functions for different values of a continuous covariate, we always choose the variation from the minimum to the maximum of the observed value.

\begin{figure}[!]
\centering
\includegraphics[width=1\linewidth]{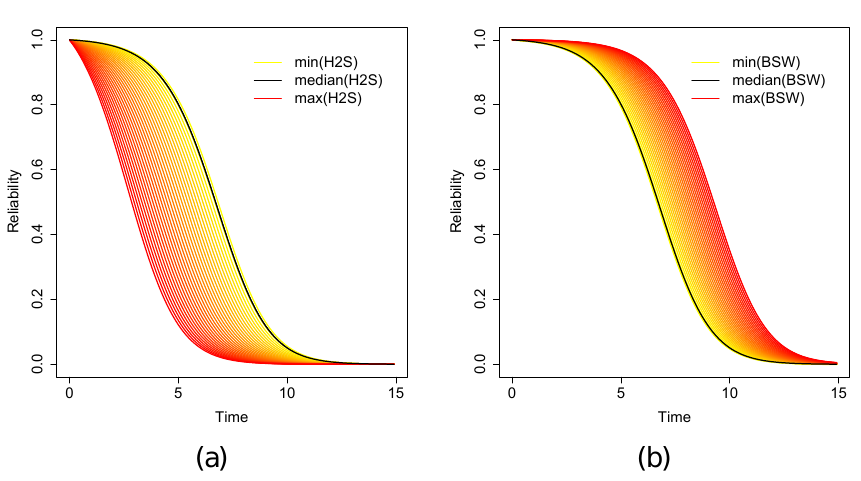}
\caption{(a) Reliability function with variation in the value of the H2S variable. (b) Reliability function with variation in the value of the BSW variable.}
\label{fig_conf_fluxo}
\end{figure}

In order to illustrate what is the effect on the reliability function for an increase in the amount of H2S or BSW, we exhibit in Figure \ref{fig_conf_fluxo} (a) several reliability curves for different values of the H2S variable. We note that, with the increase in the value of the H2S variable, the reliability function shows a faster decreasing behavior, that is, the higher the concentration of H2S, the lower the reliability of DHSVs. It is known that the concentration of H2S is associated with failures of metallic components in the oil and gas exploration industry. This can be seen, for instance, in \cite{veritas2009recommended}, which presents a summary of common threats to corrosion, with some of them involving H2S; and \cite{iso2003iso}, which gives recommendations for material selection when H2S is present.

In Figure \ref{fig_conf_fluxo} (b), we show the variation of the BSW variable being reflected in the reliability function. Observe  that  alterations  in  the  BSW  value  changed  the  reliability  curve, the higher the concentration of BSW, the greater the reliability.

\begin{figure}[!h]
\centering
\includegraphics[width=1\linewidth]{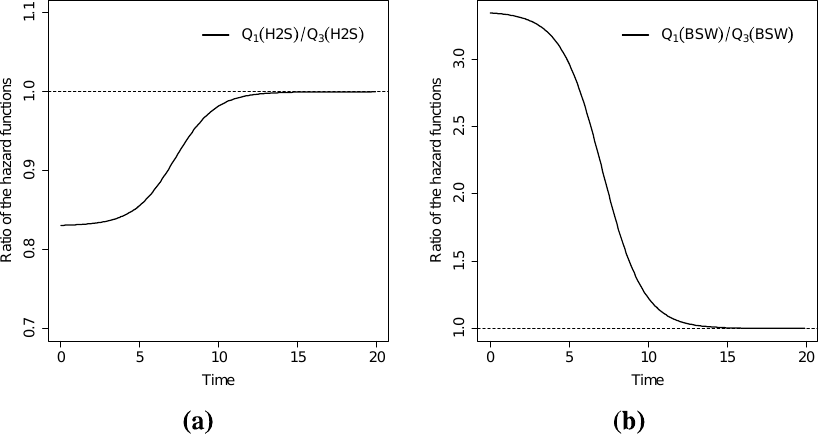}
\caption{(a) Ratio of hazard functions between the first and third quartiles of the H2S variable. (b) Ratio of hazard functions between the first and third quartiles of the BSW variable.}
\label{razao_risco_fluxo}
\end{figure}

In Figure \ref{razao_risco_fluxo} (a), we present the HR curve for the first and third quartiles of the H2S variable. We note that before 10 years of age the HR is less than one, so the risk of valve failure is greater when the H2S variable assumes the value of the third quartile; but after 10 years the HR is approximately one, and, therefore, we have approximately equal hazards. In Figure \ref{razao_risco_fluxo} (b), we show the HR for the first and third quartiles of the BSW variable.
From this plot, we observe that the HR is greater than one up to 14 years old, therefore, in this period the valve has a higher risk of failure when the BSW variable takes on the value of the first quartile; after the first 14 years the HR is approximately one, so the risk of failure is approximately equal in both groups. We highlight here a different result than the one obtained if we had considered the Cox model, since the ratio of hazard functions in the Cox model would be constant throughout the time.

\begin{figure}[!]
\centering
\includegraphics[width=1\linewidth]{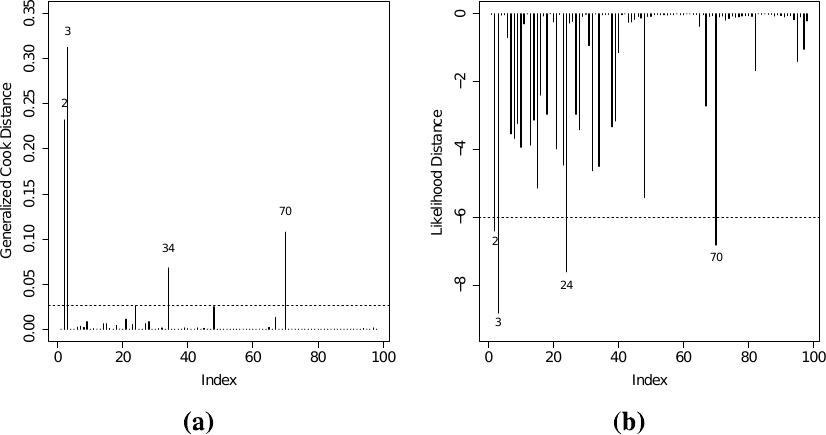}
\caption{(a) Generalized Cook's distance. (b) Likelihood distance, considering the GTDL model fitted to the flow group data.}
\label{cooks_fluxo}
\end{figure}

In order to check for the presence of influential observations, we calculated the GD and LD measures. The obtained results are presented in Figure \ref{cooks_fluxo}, from which we can see the existence of four influential observations according to the Cook's distance - cases $2,3,34$ and $70$; while from the LD, we also observe four influential observations - cases $2,3,24$ and $70$. Hence, the detected influential observations are cases $2,3,24,34$ and $70$.

We checked the impact of the detected influential cases on the
model inference. With removal of influential observations, the RC values (in percentage, \%) and p-values are displayed in Table \ref{rc_fluxo}. At the 10\% level, note that the parameters $\alpha_0$, $\beta_0$ and $\beta_1$ remained significant in all scenarios, whereas the parameter $\beta_2$ was only significant in two scenarios (specifically, when excluding the observation 24, and when excluding all influential observations). Therefore, the effect of time and the H2S variable were significant to explain the time-to-failure, while the BSW variable became non-significant in some scenarios. The RC of the parameter $\alpha$ is the largest when case 2 is excluded (RC of 20.4872\%), while for the parameters $\beta_0$ and $\beta_1$ the RCs are the largest when case 3 is removed (RC of 11.1820\% and 13.3283\%, respectively), and, finally, for the parameter $\beta_2$ the RC is the largest when case 24 is removed (RC of 48.0736\%). When all influential observations are excluded, we find that the RCs are less than 6\%, indicating little change in point estimates.

\begin{table}[!h]
\centering
\caption{The RC values (in \%) for the MLEs and SEs, in addition to the p-values, considering the deleted observations.}
\resizebox{\linewidth}{!}{
\setlength{\tabcolsep}{3pt}
\begin{tabular}{llcccc}
  \hline
Deleted case &  & $\hat{\alpha}_0$ & $\hat{\beta}_0$ & $\hat{\beta}_1$ & $\hat{\beta}_2$ \\
  \hline
   \{2\} & $\mbox{RC}_{\nu_{j(i)}}$ & 20.4872 & 9.3523 & 4.0732 & 21.7665 \\
        & $\mbox{RC}_{\mbox{SE}(\nu_{j(i)})}$ & 27.0728 & 20.3730 & 9.6097 & 4.5475 \\
        & p-value   & 0.0002     & $<$0.0001 & $<$0.0001 & 0.1718 \\
  \{3\} & $\mbox{RC}_{\nu_{j(i)}}$ &10.7815 & 11.1820 & 13.3283 & 8.0404 \\
        & $\mbox{RC}_{\mbox{SE}(\nu_{j(i)})}$ &19.7344 & 23.3812 & 15.1467 & 5.2880 \\
        & p-value   &  0.0003     & $<$0.0001 & $<$0.0001 & 0.1453 \\
 \{24\} & $\mbox{RC}_{\nu_{j(i)}}$ &5.8829 & 3.1193 & 8.2916 & 48.0736 \\
        & $\mbox{RC}_{\mbox{SE}(\nu_{j(i)})}$ &1.2191 & 3.8244 & 9.0114 & 20.5358 \\
        & p-value   &  $<$0.0001    & $<$0.0001 & $<$0.0001 & 0.0405 \\
 \{34\} & $\mbox{RC}_{\nu_{j(i)}}$ &13.2155 & 5.0062 & 1.2822 & 10.6848 \\
        & $\mbox{RC}_{\mbox{SE}(\nu_{j(i)})}$ &32.4453 & 20.6371 & 7.0519 & 3.4001 \\
        &  p-value   &  0.0008     & $<$0.0001 & $<$0.0001 & 0.1232 \\
\{70\} & $\mbox{RC}_{\nu_{j(i)}}$ &2.6705 & 6.6477 & 8.6163 & 13.3373 \\
              & $\mbox{RC}_{\mbox{SE}(\nu_{j(i)})}$ &5.9642 & 12.1104 & 8.4947 & 2.5371 \\
              & p-value   & 0.0001 & $<$0.0001 & $<$0.0001 & 0.1588 \\
\{2,3,24,34,70\} & $\mbox{RC}_{\nu_{j(i)}}$ &1.1559 & 1.0171 & 1.5947 & 5.7544 \\
              & $\mbox{RC}_{\mbox{SE}(\nu_{j(i)})}$ &1.9254 & 0.0629 & 0.0262 & 1.2419 \\
              & p-value   & 0.0001 & $<$0.0001 & $<$0.0001 & 0.0816 \\

\hline
\end{tabular} }
\label{rc_fluxo}
\end{table}

\subsection{Adjustment to the valve characteristics group}

In the structure ${\bm x}^\top\bm{\beta}$ we identified only the variable Family as statistically significant, whereas in the structure $\mathbf{x_*}^{\top}\bm{\alpha}$ we found that the variables PC and Mfr. are meaningful. Therefore, the effect of time is different for each level of these two variables. The hypothesis test for the frailty distribution's parameter resulted in the rejection of the null hypothesis. Hence, the GTDL gamma frailty model is the one that best fits these data. The obtained MLEs are shown in Table \ref{ajuste_valvula}. From the fact that the frailty parameter is significant, there is evidence that important variables were not included in the modeling, so indicating that the variables PC, Mfr. and Family are not the only ones that impact the failure time of the valves.

\begin{table}[!h]
\centering
\caption{Estimation results of the GTDL gamma frailty model fitted to the valve characteristics group.}
\resizebox{\linewidth}{!}{
\setlength{\tabcolsep}{3pt}
\begin{tabular}{l|cccc}
  \hline
         Parameter         & MLE     & SE  &  90\% CI \\
  \hline
  $\alpha_0$                & -5.3280 & 0.7101 & (-6.4961; -4.1599) \\
  $\alpha_1$ (7,500)        & 1.9430 & 0.7919  & (0.6403; 3.2457)  \\
  $\alpha_2$ (10,000)       & 0.8336 & 0.3071  & (0.3284; 1.3388)  \\
  $\alpha_3$ (Mfr. B)       & 5.6018 & 0.7466  & (4.3736; 6.8300)  \\
  $\alpha_4$ (Mfr. A)       & 5.8969 & 0.7148  & (4.7211; 7.0727)  \\
  $\beta_0$                 & -6.1317& 1.1002  & (-7.9415; -4.3219) \\
  $\beta_1$  (family B)     & 0.8631 & 1.2455  & (-1.1857; 2.9119) \\
  $\beta_2$  (Others)       & 5.8098 & 1.6673  & (3.0671; 8.5525)  \\
  $\theta$                  &12.3951 & 3.5735  & (6.5166; 18.2735) \\
   \hline
\end{tabular} }
\label{ajuste_valvula}
\end{table}

It is worth mentioning that when the Manufacturer is the ``Others'' class, the GTDL with gamma frailty term assumes a cure fraction, regardless of what the other explanatory variables are, since in this case $\mathbf{x_*}^{\top}\bm{\alpha}<0$, with the cure fraction value close to one. From a descriptive analysis, we identified that only two out of  30 observations in the ``Others'' class were failure times, occurring before one year of operation.

\begin{figure}[h]
\centering
\includegraphics[width=0.5\linewidth]{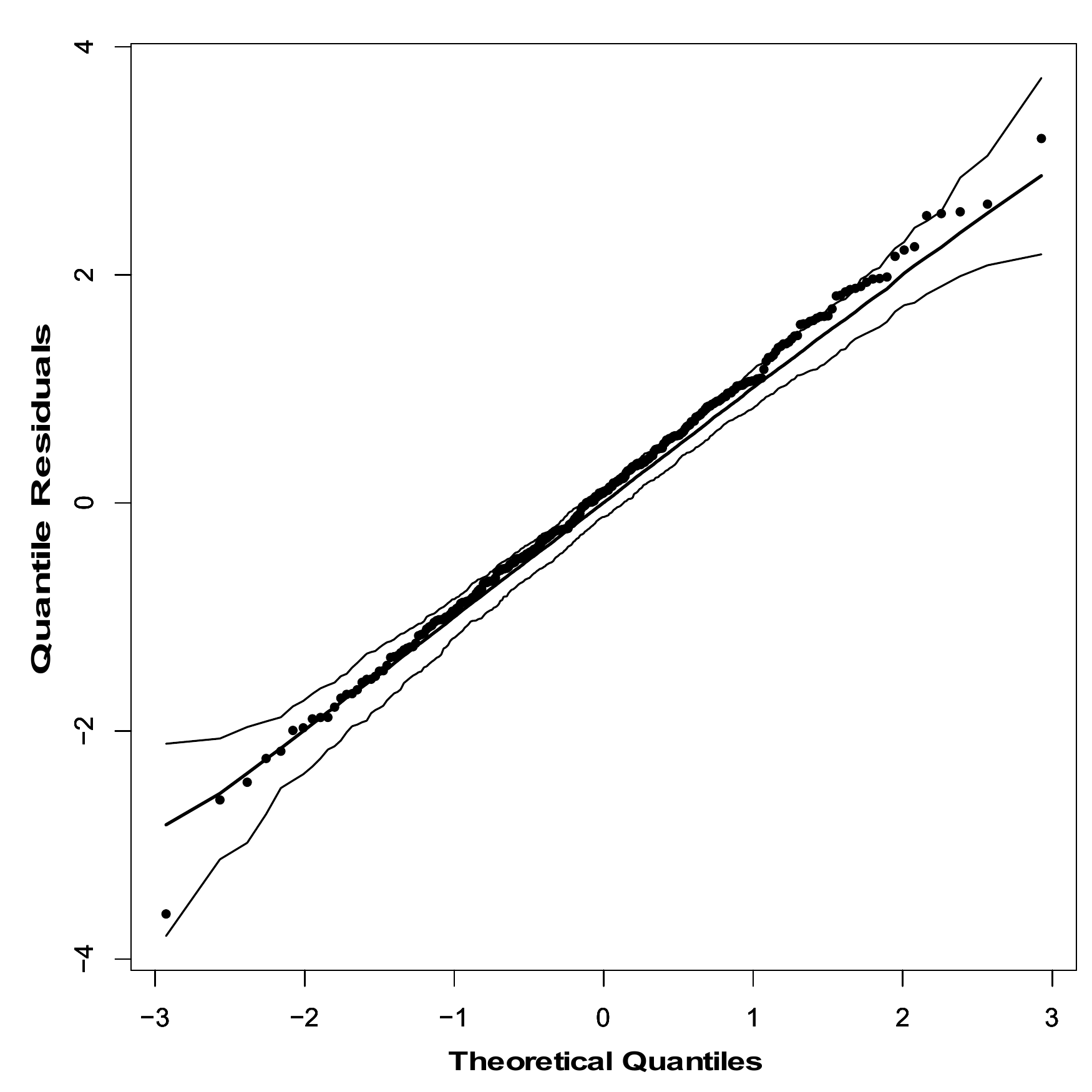}
\caption{QQ-plot with envelope of 95\% for the RQ residuals of the adjustment to the valve characteristics group.}
\label{residuo_valvula}
\end{figure}

In Figure \ref{residuo_valvula}, we present the QQ-plot of the RQ residuals. Again, we observed a good agreement between the residuals and the standard normal distribution. Therefore, we can conclude that there was a good adjustment of the GTDL gamma frailty model to the data.

\begin{figure}[!h]
\centering
\includegraphics[width=1\linewidth]{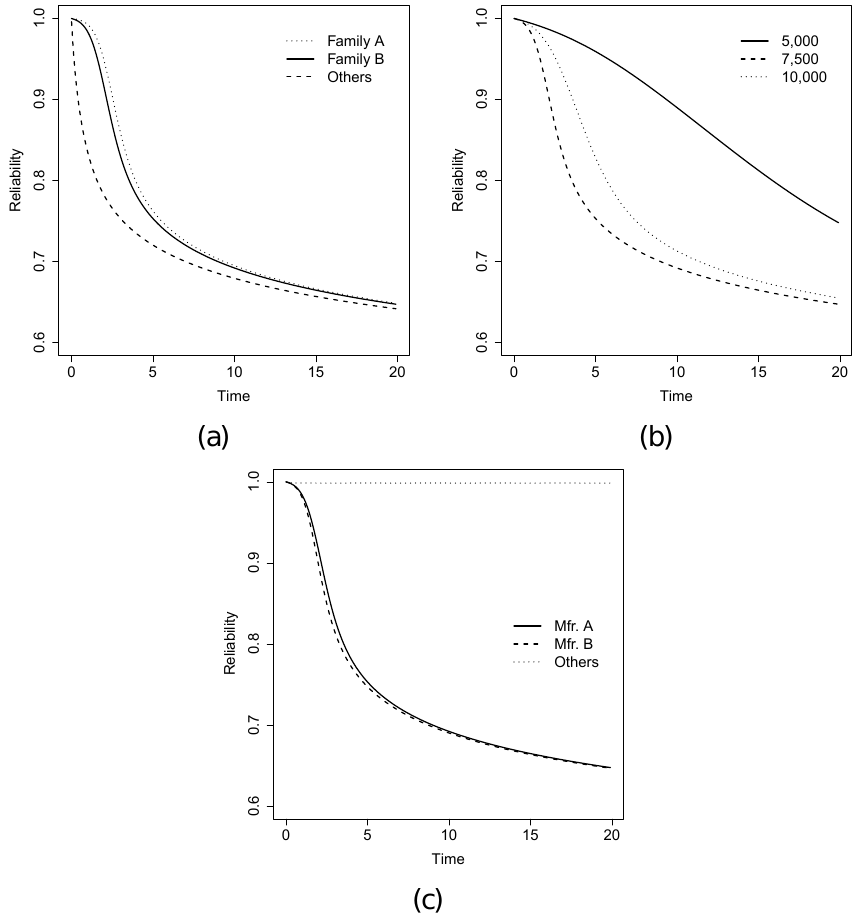}
\caption{(a) Reliability function for the Family variable, considering PC equal to ``7,500'' and ``Mfr. B'' for the Manufacturer. (b) Reliability function for the PC variable, adopting ``family B'' for Family and ``Mfr. B'' for Manufacturer. (c) Reliability function for the Mfr. variable, assuming ``family B'' for Family and ``7,500'' for PC.}
\label{fig_conf_valvula}
\end{figure}

Figure \ref{fig_conf_valvula} shows the reliability functions for the variables Family, PC and Mfr.. Note that there is a little difference between the reliability curves of the ``family A'' and ``family B'' classes;
the same occurs with the ``Mfr. A'' and ``Mfr. B'' classes' curves.
By analyzing the curves of the PC variable, we see that the lowest reliability is for PC equal to ``7,500'', while the highest one is for PC equal to ``5,000''.

The closeness between the reliability curves of the ``family B'' and ``family A'' levels is justified by analyzing the confidence interval of the ``family B'' Family, since this level is not statistically different from the ``family A'' reference level. From this, we can conclude that the failure times showed no significant difference in relation to these two levels of the Family variable. The same conclusion can be made for the reliability curves considering the PC ``7,500'' and ``10,000'', and also the Manufacturers ``Mfr. A'' and ``Mfr. B''. But for that, it was necessary to change the reference classes and refit the model.

Due to the architecture imposed for safety valves in deep water wells used by Petrobras, most of its valves have nitrogen chambers, which is a technology that mitigates the pressure sensitivity of the well and ensures, through a surface calibration for the individual condition of each well, the opening and closing pressures of a particular specification. In other words, the pressure class envisaged in the analysis, consists of a control variable and easy handling for new DHSVs. From the fact that the calibration is feasible and a low-cost action, with a relative impact on reducing the risks of that component, it is advisable to use the PC equal to 5,000 psi.

\begin{figure}[!]
\centering
\includegraphics[width=1\linewidth]{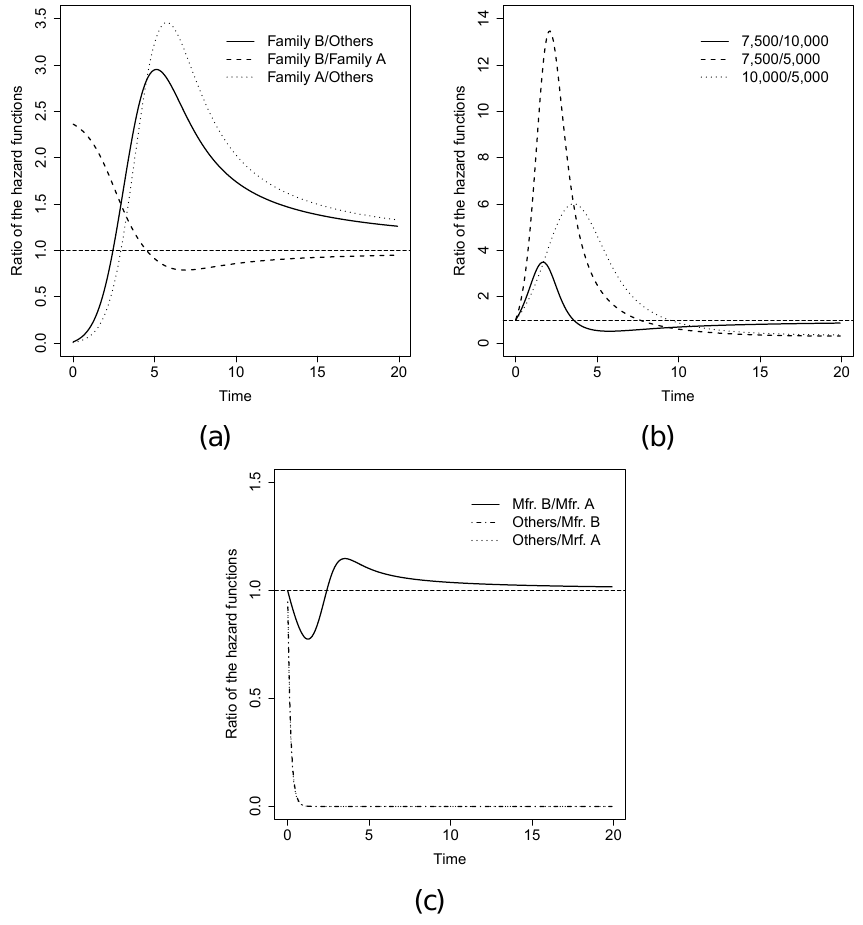}
\caption{(a) Ratio of hazard functions of the Family variable, adopting PC equal to ``10,000'' and Manufacturer ``Mfr. B''. (b) Ratio of hazard functions of PC, adopting the ``family B'' class for Family and Manufacturer ``Mfr. B''. (c) Ratio of hazard functions of Manufacturer, adopting the ``family B'' class for Family and PC equal to ``7,500''.}
	\label{razao_risco_valvula}
\end{figure}

Figure \ref{razao_risco_valvula} presents an analysis of the behavior of the hazard function using the GTDL gamma frailty model fitted to the data. From the HR of the Family variable shown in Figure \ref{razao_risco_valvula} (a), we observe that the ratios between ``family B'' and ``Others'', ``family A'' and ``Others'' start below one, so indicating that the ``Others'' family has a higher risk; however, after approximately 3 years this relationship is reversed. For the ratio between ``family B'' and ``family A'', we see that it is greater than one until approximately 5 years, i.e., at the beginning ``family B'' shows a higher risk of failure compared to ``family A''; but after 5 years it is ``family A'' that exhibits a greater risk of failure.

From the HRs for the PC variable displayed in Figure \ref{razao_risco_valvula} (b), we observe that, initially, the PC of ``10,000'' shows more risk of failure than the PC of ``5,000'', and the PC of ``7,500'' is more at risk of failure than PC of ``5,000'' and ``10,000''; nevertheless, these relationships are reversed over time. It is worth noting that the risk of failure of PC ``7,500'' reaches approximately 14 times the risk of failure of PC ``5,000''. When the comparison is made with PC ``10,000'', the risk of PC ``7,500'' is even approximately 6 times.

Finally, Figure \ref{razao_risco_valvula} (c) exhibits the HRs for the Mfr. variable, from which we note that the ratios between ``Others'' and ``Mfr. A'', ``Others'' and ``Mfr. B'' are always below one, so indicating that the risk of failure is greater for the Manufacturers ``Mfr. B'' and ``Mfr. A''. When comparing ``Mfr. B'' and ``Mfr. A'', we observe that, initially, the ratio is less than one until the age of approximately 3 years, thus indicating that ``Mfr. A'' has a higher risk of failure than ``Mfr. B''. However, after 3 years the relationship is inverted and maintained over time.

\begin{figure}[!]
\centering
\includegraphics[width=1\linewidth]{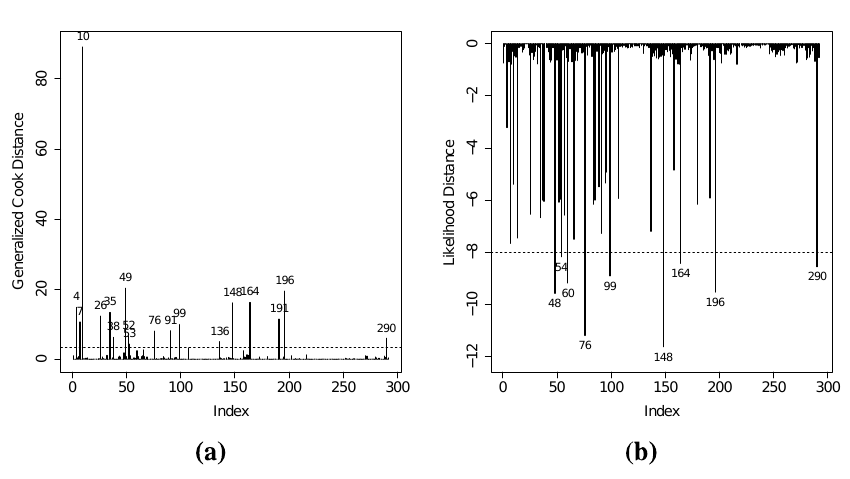}
\caption{(a) Generalized Cook's distance. (b) Likelihood distance, considering the GTDL gamma frailty model fitted to the valve group data.}
\label{cooks_valvula}
\end{figure}

The GD measure identified 18 influential observations, while the LD indicated 9 influential observations; these indications can be seen in Figure \ref{cooks_valvula}. The cases $76, 99, 148, 164, 196$ and $290$ were detected by both metrics.

In Table \ref{rc_valvula}, we observe that the MLEs of the parameters $\alpha_0$, $\alpha_1$, $\alpha_2$, $\alpha_3$, $\alpha_4$, $\beta_0$, $\beta_2$ and $\theta$ are always statistically significant at the 10\% level,
while the MLE of $\beta_1$ is only meaningful when removing all influential observations. We also note a big change in the parameter estimates when all the influential observations are deleted, e.g., the estimates of the parameters $\alpha_1$, $\alpha_2 $, $\alpha_3 $, $\alpha_4 $, $\beta_1$ and $\theta$, change from 1.9430, 0.8336, 5.6018, 5.8969, 0.8631 and 12.3951, to 12.2712, 10.4262, -27.2694, -19.7312, 34.6029 and 64.3544, respectively. This makes us believe that the reason for this alteration is that all removed observations are due to failure times.

\begin{table}[!h]
\centering
\caption{The RC values (in \%) for the MLEs and SEs, in addition to the p-values, considering the deleted observations.}
\resizebox{\linewidth}{!}{
\setlength{\tabcolsep}{3pt}
\begin{tabular}{llcccccccccc}
  \hline
Deleted case & & $\hat{\alpha}_0$ & $\hat{\alpha}_1$ & $\hat{\alpha}_2$ & $\hat{\alpha}_3$ & $\hat{\alpha}_4$ & $\hat{\beta}_0$& $\hat{\beta}_1$& $\hat{\beta}_2$ & $\hat{\theta}$ \\
  \hline
\{4\}     & $\mbox{RC}_{\nu_{j(i)}}$            &30.4017 & 1.4089 & 0.6052 & 29.2493 & 27.4773 & 0.0645 & 5.5488 & 15.4158 & 0.1140 \\
          & $\mbox{RC}_{\mbox{SE}(\nu_{j(i)})}$ &87.8263 & 2.0345 & 1.4861 & 81.4457 & 87.2687 & 0.3251 & 0.8465 & 1.4755 & 6.7392 \\
          & p-value                             &0.0054 & 0.0177 & 0.0056 & 0.0034 & 0.0014 & $<$0.0001 & 0.4607 & 0.0028 & 0.0011 \\
\{7\}     & $\mbox{RC}_{\nu_{j(i)}}$            &0.4452 & 0.3912 & 1.5539 & 0.0453 & 0.2204 & 1.2104 & 10.1491 & 0.5057 & 7.3346 \\
          & $\mbox{RC}_{\mbox{SE}(\nu_{j(i)})}$ &37.6123 & 0.3408 & 1.7897 & 34.5184 & 38.0240 & 6.1609 & 5.1090 & 4.3746 & 7.1643 \\
          & p-value                             &$<$0.0001 & 0.0134 & 0.0087 & $<$0.0001 & $<$0.0001 & $<$0.0001 & 0.5536 & 0.0009 & 0.0005 \\
\{10\}    & $\mbox{RC}_{\nu_{j(i)}}$            &83.5834 & 0.0054 & 0.8636 & 79.3918 & 75.5341 & 0.1580 & 1.6602 & 4.8208 & 0.6080 \\
          & $\mbox{RC}_{\mbox{SE}(\nu_{j(i)})}$ &52.8768 & 0.4091 & 0.3253 & 47.7292 & 50.6990 & 0.2150 & 0.5921 & 2.1095 & 2.3203 \\
          & p-value                             &$<$0.0001 & 0.0145 & 0.0060 & $<$0.0001 & $<$0.0001 & $<$0.0001 & 0.4785 & 0.0012 & 0.0006 \\
\{26\}    & $\mbox{RC}_{\nu_{j(i)}}$            & 0.7032 & 0.1040 & 4.3077 & 0.5007 & 0.5018 & 0.8333 & 7.7694 & 1.4911 & 7.8598 \\
          & $\mbox{RC}_{\mbox{SE}(\nu_{j(i)})}$ & 2.7794 & 0.3888 & 7.7963 & 2.8082 & 2.6908 & 2.6268 & 1.4871 & 2.6859 & 8.3664 \\
          & p-value                             & $<$0.0001 & 0.0144 & 0.0086 & $<$0.0001 & $<$0.0001 & $<$0.0001 & 0.4618 & 0.0006 & 0.0006 \\
\{35\}    &$\mbox{RC}_{\nu_{j(i)}}$             & 0.6100 & 0.4441 & 5.0868 & 0.3712 & 0.4040 & 0.9923 & 7.4050 & 1.7257 & 8.1467 \\
          &$\mbox{RC}_{\mbox{SE}(\nu_{j(i)})}$  & 2.5757 & 0.3541 & 7.5707 & 2.6490 & 2.4911 & 2.7222 & 1.6881 & 2.7972 & 8.3420 \\
          &p-value                              & $<$0.0001 & 0.0141 & 0.0080 & $<$0.0001 & $<$0.0001 & $<$0.0001 & 0.4642 & 0.0006 & 0.0005 \\
\{38\}    & $\mbox{RC}_{\nu_{j(i)}}$            & 1.2109 & 1.6831 & 1.1012 & 1.2304 & 1.0573 & 0.2682 & 7.6990 & 0.0978 & 5.6408 \\
          & $\mbox{RC}_{\mbox{SE}(\nu_{j(i)})}$ & 3.7460 & 0.2957 & 5.3532 & 3.4772 & 3.6495 & 1.2140 & 0.2522 & 1.4383 & 7.5534 \\
          & p-value                             & $<$0.0001 & 0.0162 & 0.0108 & $<$0.0001 & $<$0.0001 & $<$0.0001 & 0.4566 & 0.0006 & 0.0007 \\
\{48\}    &$\mbox{RC}_{\nu_{j(i)}}$             & 0.6206 & 10.6976 & 26.2448 & 4.0939 & 0.9822 & 5.5191 & 48.3224 & 5.4630 & 1.1571 \\
          &$\mbox{RC}_{\mbox{SE}(\nu_{j(i)})}$  & 14.3356 & 3.3399 & 7.6441 & 13.8904 & 13.0563 & 5.4882 & 1.0541 & 3.4839 & 2.9844 \\
          &p-value                              & $<$0.0001 & 0.0086 & 0.0015 & $<$0.0001 & $<$0.0001 & $<$0.0001 & 0.3091 & 0.0004 & 0.0009 \\
\{49\}    &$\mbox{RC}_{\nu_{j(i)}}$             & 34.6316 & 2.1279 & 0.1467 & 33.3239 & 31.3236 & 0.0332 & 6.3699 & 18.1310 & 1.2663 \\
          &$\mbox{RC}_{\mbox{SE}(\nu_{j(i)})}$  & 107.5781 & 2.6918 & 1.8539 & 99.9393 & 106.9684 & 0.5716 & 1.1461 & 2.8460 & 6.0849 \\
          &p-value                              & 0.0181 & 0.0194 & 0.0056 & 0.0123 & 0.0062 & $<$0.0001 & 0.4559 & 0.0033 & 0.0012 \\
\{52\}    &$\mbox{RC}_{\nu_{j(i)}}$             & 1.1845 & 1.5895 & 0.7249 & 1.1900 & 1.0265 & 0.1917 & 7.8443 & 0.0108 & 5.8111 \\
          &$\mbox{RC}_{\mbox{SE}(\nu_{j(i)})}$  & 3.7452 & 0.3120 & 5.6929 & 3.4954 & 3.6449 & 1.3445 & 0.3291 & 1.5451 & 7.6575 \\
          &p-value                              & $<$0.0001 & 0.0161 & 0.0108 & $<$0.0001 & $<$0.0001 & $<$0.0001 & 0.4563 & 0.0006 & 0.0007 \\
\{53\}    &$\mbox{RC}_{\nu_{j(i)}}$             & 1.3247 & 2.0415 & 2.9656 & 1.4077 & 1.1947 & 0.6467 & 6.5073 & 0.6328 & 4.7346 \\
          &$\mbox{RC}_{\mbox{SE}(\nu_{j(i)})}$  & 3.6747 & 0.2106 & 3.4059 & 3.3185 & 3.6000 & 0.5177 & 0.0978 & 0.8744 & 6.9317 \\
          &p-value                              & $<$0.0001 & 0.0165 & 0.0109 & $<$0.0001 & $<$0.0001 & $<$0.0001 & 0.4600 & 0.0006 & 0.0007 \\
\{54\}    &$\mbox{RC}_{\nu_{j(i)}}$             & 2.0334 & 9.6956 & 1.0775 & 2.6364 & 1.8700 & 0.2125 & 38.8591 & 1.3556 & 1.6908 \\
          &$\mbox{RC}_{\mbox{SE}(\nu_{j(i)})}$  & 2.9805 & 0.4837 & 0.2554 & 2.3465 & 2.9503 & 0.2574 & 2.8941 & 1.3314 & 1.0405 \\
          &p-value                              & $<$0.0001 & 0.0074 & 0.0059 & $<$0.0001 & $<$0.0001 & $<$0.0001 & 0.6805 & 0.0005 & 0.0005 \\
\{60\}    &$\mbox{RC}_{\nu_{j(i)}}$             & 3.2507 & 13.4936 & 5.3774 & 4.1729 & 3.0445 & 1.0967 & 49.4690 & 2.9123 & 2.7842 \\
          &$\mbox{RC}_{\mbox{SE}(\nu_{j(i)})}$  & 4.8711 & 0.5454 & 2.1287 & 3.8186 & 4.8151 & 1.4059 & 4.0903 & 2.6415 & 1.3689 \\
          &p-value                              & $<$0.0001 & 0.0056 & 0.0051 & $<$0.0001 & $<$0.0001 & $<$0.0001 & 0.7366 & 0.0005 & 0.0004 \\
\{76\}    &$\mbox{RC}_{\nu_{j(i)}}$             & 5.3734 & 19.7508 & 13.7541 & 6.8757 & 5.1136 & 2.8191 & 65.7080 & 5.8936 & 5.3690 \\
          &$\mbox{RC}_{\mbox{SE}(\nu_{j(i)})}$  & 7.8656 & 0.9865 & 7.5837 & 6.0300 & 7.7674 & 3.8681 & 6.5047 & 5.4044 & 2.9683 \\
          &p-value                              & $<$0.0001 & 0.0036 & 0.0041 & $<$0.0001 & $<$0.0001 & $<$0.0001 & 0.8234 & 0.0005 & 0.0004 \\
\{91\}    &$\mbox{RC}_{\nu_{j(i)}}$             & 0.4816 & 0.3599 & 1.3006 & 0.1340 & 0.9395 & 3.3248 & 24.9909 & 2.8898 & 6.3258 \\
          &$\mbox{RC}_{\mbox{SE}(\nu_{j(i)})}$  & 10.3862 & 0.3317 & 1.1331 & 9.6004 & 10.5617 & 1.5611 & 1.8674 & 1.9503 & 6.6533 \\
          &p-value                              & $<$0.0001 & 0.0135 & 0.0081 & $<$0.0001 & $<$0.0001 & $<$0.0001 & 0.6099 & 0.0009 & 0.0005 \\
\{99\}    &$\mbox{RC}_{\nu_{j(i)}}$             & 5.3029 & 2.7492 & 1.6940 & 5.3210 & 4.8801 & 0.3085 & 4.3383 & 3.6137 & 6.7477 \\
          &$\mbox{RC}_{\mbox{SE}(\nu_{j(i)})}$  & 6.6174 & 1.4876 & 1.7026 & 5.7547 & 6.5723 & 1.0905 & 1.0761 & 3.9151 & 6.1229 \\
          &p-value                              & $<$0.0001 & 0.0105 & 0.0066 & $<$0.0001 & $<$0.0001 & $<$0.0001 & 0.5119 & 0.0005 & 0.0005 \\
\{136\}   &$\mbox{RC}_{\nu_{j(i)}}$             & 12.2470 & 1.6784 & 0.2632 & 11.9606 & 11.0912 & 0.0027 & 5.1213 & 15.6160 & 1.1150 \\
          &$\mbox{RC}_{\mbox{SE}(\nu_{j(i)})}$  & 21.6595 & 2.0991 & 1.5249 & 19.1983 & 21.5928 & 0.3857 & 0.9612 & 4.2859 & 4.5029 \\
          &p-value                              & $<$0.0001 & 0.0181 & 0.0057 & $<$0.0001 & $<$0.0001 & $<$0.0001 & 0.4620 & 0.0021 & 0.0010 \\
\{148\}   &$\mbox{RC}_{\nu_{j(i)}}$             & 0.1586 & 7.5269 & 17.3875 & 0.4677 & 2.1281 & 18.1920 & 110.3501 & 19.2958 & 2.6873 \\
          &$\mbox{RC}_{\mbox{SE}(\nu_{j(i)})}$  & 12.7337 & 0.0151 & 6.5753 & 11.2412 & 12.8310 & 20.1945 & 12.3395 & 9.8633 & 2.5618 \\
          &p-value                              & $<$0.0001 & 0.0083 & 0.0028 & $<$0.0001 & $<$0.0001 & $<$0.0001 & 0.1944 & 0.0002 & 0.0005 \\
\{164\}   &$\mbox{RC}_{\nu_{j(i)}}$             & 18.0517 & 2.1956 & 1.9246 & 17.3682 & 16.4088 & 0.3663 & 2.5953 & 9.3629 & 7.0554 \\
          &$\mbox{RC}_{\mbox{SE}(\nu_{j(i)})}$  & 21.8948 & 0.8955 & 2.8491 & 18.3403 & 21.5007 & 1.4035 & 2.6764 & 13.1805 & 6.4181 \\
          &p-value                              & $<$0.0001 & 0.0114 & 0.0071 & $<$0.0001 & $<$0.0001 & $<$0.0001 & 0.5109 & 0.0008 & 0.0005 \\
\{191\}   &$\mbox{RC}_{\nu_{j(i)}}$             & 0.6231 & 2.7870 & 4.1934 & 0.4883 & 0.6648 & 2.1828 & 22.6979 & 1.6876 & 7.5923 \\
          &$\mbox{RC}_{\mbox{SE}(\nu_{j(i)})}$  & 3.8902 & 1.1116 & 11.5756 & 3.8539 & 3.8108 & 1.1182 & 0.9392 & 1.7019 & 7.7862 \\
          &p-value                              & $<$0.0001 & 0.0126 & 0.0112 & $<$0.0001 & $<$0.0001 & $<$0.0001 & 0.5956 & 0.0008 & 0.0005 \\
\{196\}   &$\mbox{RC}_{\nu_{j(i)}}$             & 21.9517 & 2.1653 & 0.7409 & 21.4201 & 19.9186 & 0.1242 & 7.4741 & 10.8301 & 7.2652 \\
          &$\mbox{RC}_{\mbox{SE}(\nu_{j(i)})}$  & 24.3956 & 0.8451 & 3.5498 & 20.6982 & 24.0475 & 1.3982 & 2.3772 & 14.0496 & 6.2292 \\
          &p-value                              & $<$0.0001 & 0.0115 & 0.0083 & $<$0.0001 & $<$0.0001 & $<$0.0001 & 0.5311 & 0.0007 & 0.0005 \\
\{290\}   &$\mbox{RC}_{\nu_{j(i)}}$             & 0.1946 & 1.0355 & 2.3766 & 0.2919 & 1.3768 & 11.3504 & 83.0588 & 12.0761 & 0.2652 \\
          &$\mbox{RC}_{\mbox{SE}(\nu_{j(i)})}$  & 5.0417 & 0.1510 & 0.0884 & 4.7237 & 5.1739 & 15.2285 & 9.9903 & 6.8935 & 1.0658 \\
          &p-value                              & $<$0.0001 & 0.0153 & 0.0081 & $<$0.0001 & $<$0.0001 & $<$0.0001 & 0.2488 & 0.0003 & 0.0006 \\
\{All\} &$\mbox{RC}_{\nu_{j(i)}}$             & 639.0315 & 531.5499 & 1150.6768 & 586.7832 & 434.5975 & 849.2352 & 3909.2424 & 2257.1743 & 419.2168 \\
          & $\mbox{RC}_{\mbox{SE}(\nu_{j(i)})}$ & 115.8262 & 462.4476 & 1129.8854 & 245.8145 & 228.6784 & 1888.1650 & 1178.6618 & 1190.1720 & 946.7594 \\
          &p-value                              & $<$0.0001 & 0.0059 & 0.0058 & $<$0.0001 & $<$0.0001 & 0.0078 & 0.0298 & $<$0.0001 & 0.0853 \\  \hline
\end{tabular} }
\label{rc_valvula}
\end{table}

\subsection{Adjustment to the environment characteristics group}

In Step 1, the variables OU, CWT and WC are statistically significant; while in Step 2, only the intercept is relevant.  The GTDL model (without frailty) is the one that best fits these data, since we do not reject the null hypothesis ($H_0: \theta=0$) at the 10\% significance level. From the obtained MLEs displayed in Table \ref{ajuste_ambiente}, we observe that all parameters are significant at the 10\% level, except for the parameter $\beta_1$, which measures the effect of the OU class ``SB''. Thus, we can say that there is no significant difference between the OU levels ``CB'' (reference) and ``SB''.

\begin{table}[!h]
\centering
\caption{Estimation results of the GTDL model fitted to the environment characteristics group.}
\resizebox{\linewidth}{!}{
\setlength{\tabcolsep}{3pt}
\begin{tabular}{l|ccc}
  \hline
          Parameter          & MLE     & SE  &  90\% CI \\
  \hline
$\alpha_0$          &  0.1542 & 0.0277 & (0.1087; 0.1996) \\
$\beta_0$           & -5.0075 & 0.5202 & (-5.8632; -4.1518) \\
$\beta_1$ (OU-SB)   & -0.5979 & 0.5185 & (-1.4507; 0.2550) \\
$\beta_2$ (OU-ES)   &  2.1236 & 0.4910 & (1.3158; 2.9314) \\
$\beta_3$ (CWT)    &  0.0203 & 0.0067 & (0.0093; 0.0314) \\
$\beta_4$ (WC)     &  0.6172 & 0.3611 & (0.0233; 1.2111) \\
   \hline
\end{tabular} }
\label{ajuste_ambiente}
\end{table}

\begin{figure}[h]
\centering
\includegraphics[width=0.5\linewidth]{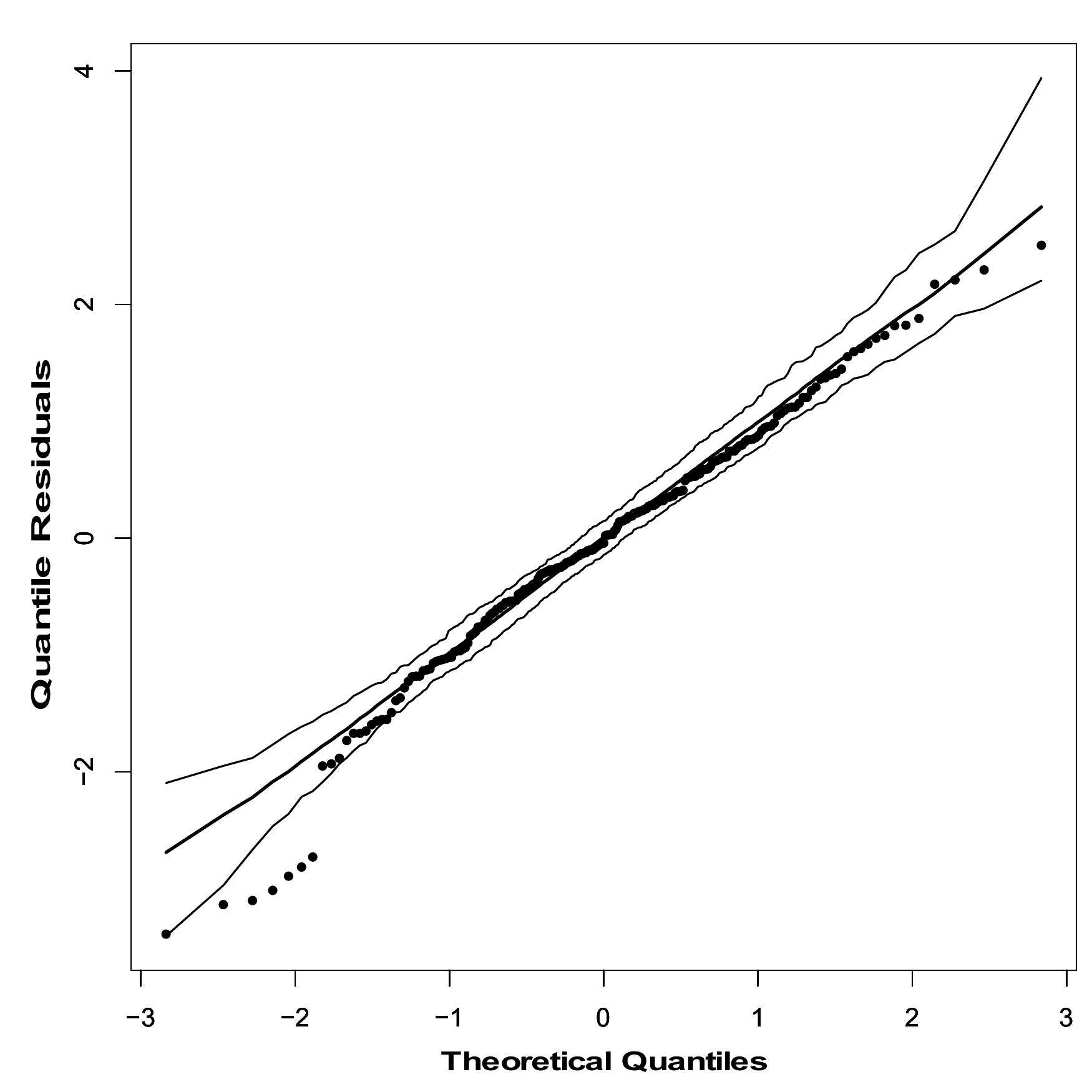}
\caption{QQ-plot with envelope of 95\% for the RQ residuals of the adjustment to the environment characteristics group.}
\label{residuo_ambiental}
\end{figure}

Figure \ref{residuo_ambiental} shows the QQ-plot of the RQ residuals. In general, we observed a good agreement between the residuals and the standard normal distribution, but we noticed a slight deviation in the lower tail.

\begin{figure}[!]
\centering
\includegraphics[width=1\linewidth]{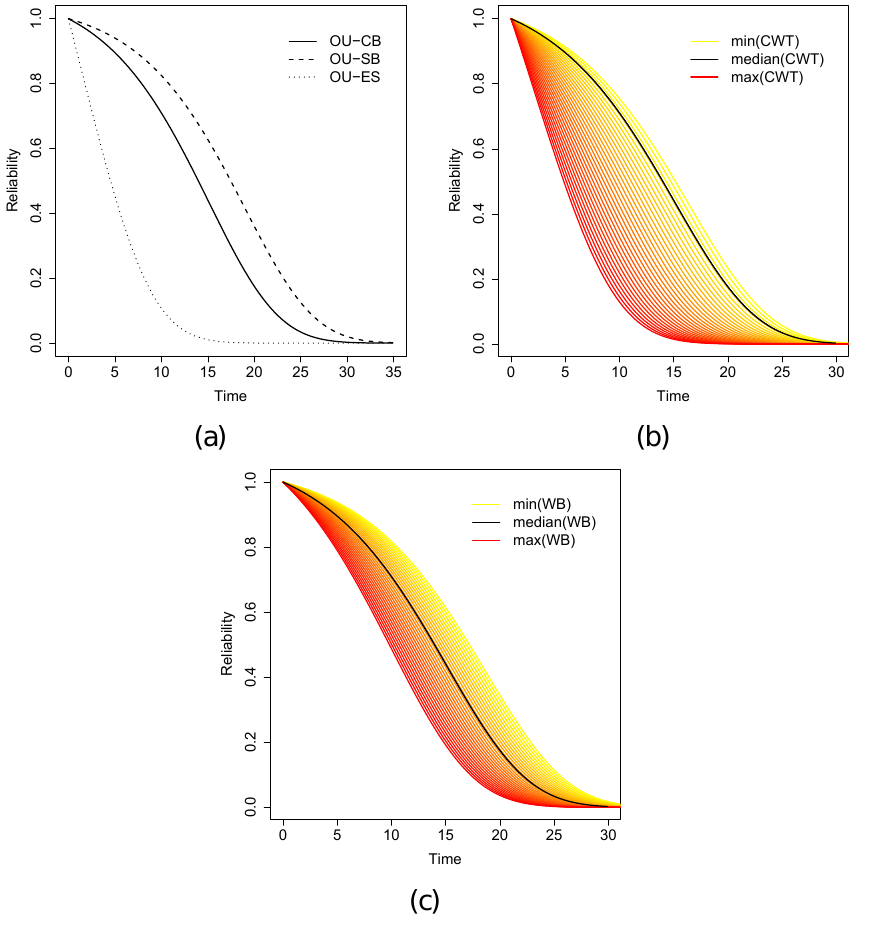}
\caption{(a) Reliability function for the operating units. (b) Reliability function with variation in the value of the CWT variable. (c) Reliability function with variation in the value of the WC variable.}
	\label{fig_conf_ambiente}
\end{figure}

From Figure \ref{fig_conf_ambiente},
we note that ``ES'' is the class with the lowest reliability among the three operating units, while ``SB'' is the one with the highest reliability. Moreover, it can be observed that the higher the value of the CWT and WC variables, the lower the reliability.

\begin{figure}[!]
\centering
\includegraphics[width=1\linewidth]{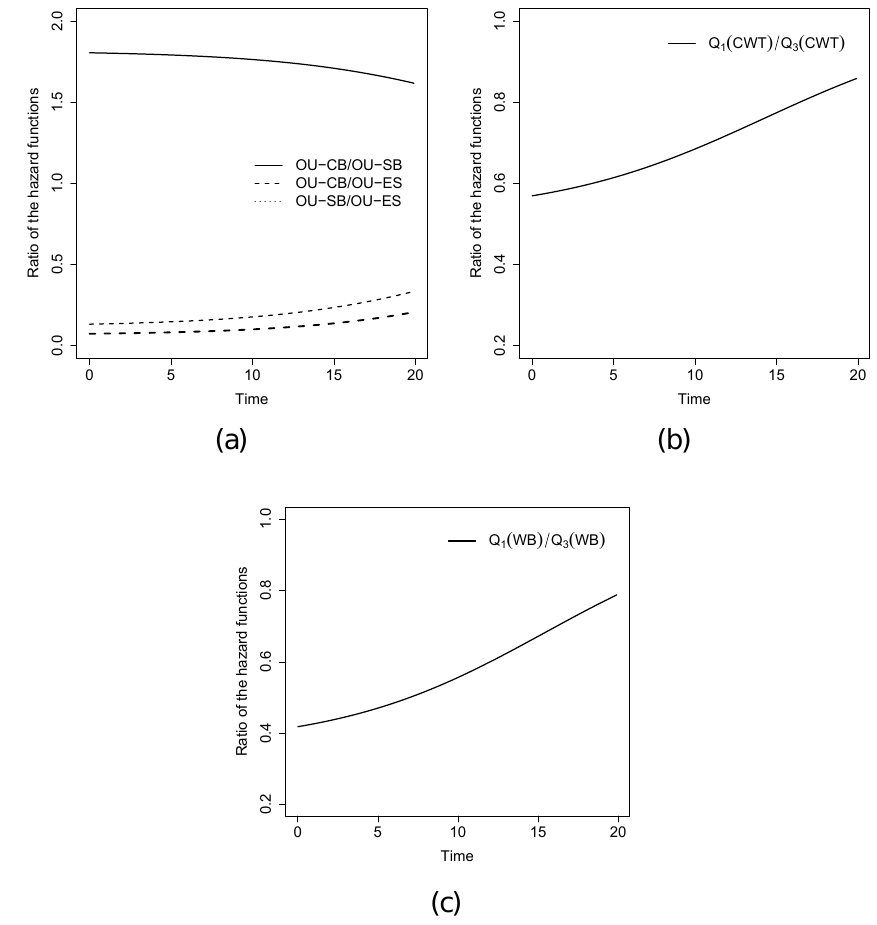}
\caption{(a) Ratio of hazard functions of the OU variable. (b) Ratio of hazard functions between the first and third quartiles of the CWT variable.
(c) Ratio of hazard functions between the first and third quartiles of the WC variable.}
	\label{razao_risco_ambiente}
\end{figure}

Figure \ref{razao_risco_ambiente} (a) shows that the HRs between ``CB'' and ``SB'' with ``ES'' are below one all the time, so indicating that the risk of valve failure is lower in the ``CB'' and ``SB'' operating units. When analyzing the HR between the ``CB'' and ``SB'', we note that it is always greater than one, thus indicating that ``SB'' has a lower risk of failure. Finally, by analyzing the HRs involving the first and third quartiles of the CWT and WC variables (Figure \ref{razao_risco_ambiente} (b) and Figure \ref{razao_risco_ambiente} (c), respectively), we observe that both are less than one, thus indicating that the increase in the value of these variables causes the risk to also increase.

\begin{figure}[!]
\centering
\includegraphics[width=1\linewidth]{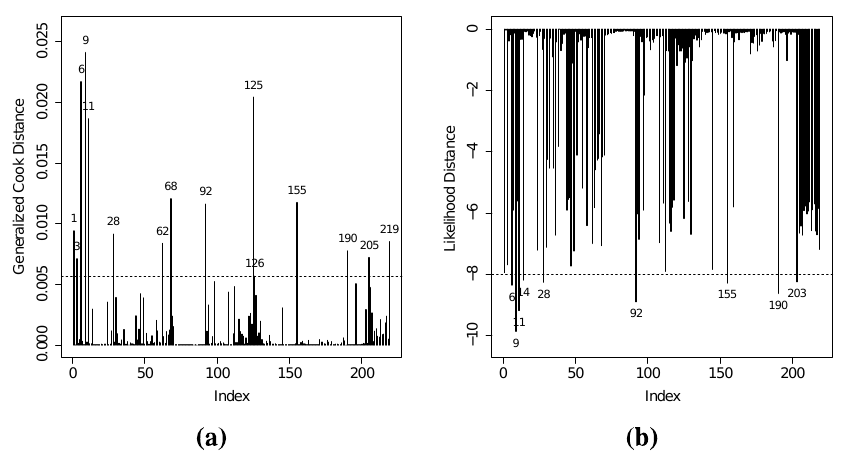}
\caption{(a) Generalized Cook's distance. (b) Likelihood distance, considering the GTDL model fitted to the environmental group data.}
\label{cooks_ambiente}
\end{figure}

Finally, Figure \ref{cooks_ambiente} exhibits the GD and LD measures, considering the GTDL model fitted to the environmental group data.
In total, 17 influential observations are detected. In Table \ref{influencia_ambiente}, we present the RCs and p-values, from which we note that the effect of time, the CWT variable and the OU class ``ES'' are significant at  the  10\% level in all arrangements. While the OU class ``SB'' is not significant in any of the scenarios, and the WC variable is significant in most of the scenarios. The RC is the largest when we remove all the influential observations, and this result is valid for all parameters. These changes range from 29.1455\% (for the parameter $\beta_0$) to 65.9195\% (for the parameter $\beta_2$).

\begin{table}[!]
\centering
\caption{The RC values (in \%) for the MLEs and SEs, in addition to the p-values, considering the deleted observations.}
\resizebox{\linewidth}{!}{
\setlength{\tabcolsep}{3pt}
\begin{tabular}{clcccccc}
  \hline
  Deleted case & & $\hat{\alpha}_0$ & $\hat{\beta}_0$ & $\hat{\beta}_1$ & $\hat{\beta}_2$ & $\hat{\beta}_3$ & $\hat{\beta}_4$\\
  \hline
   \{1\} & $\mbox{RC}_{\nu_{j(i)}}$            & 2.9254 & 2.9205 & 2.5237 & 1.0057 & 7.2610 & 10.9126 \\
         & $\mbox{RC}_{\mbox{SE}(\nu_{j(i)})}$ & 1.7930 & 3.6059 & 0.4641 & 0.3209 & 1.5571 & 1.8379 \\
         & p-value                             & $<$0.0001 & $<$0.0001 & 0.2393 & $<$0.0001 & 0.0014 & 0.0626 \\
   \{3\} & $\mbox{RC}_{\nu_{j(i)}}$            & 4.2387 & 2.4626 & 1.4264 & 0.1815 & 1.8372 & 11.9369 \\
         & $\mbox{RC}_{\mbox{SE}(\nu_{j(i)})}$ & 1.5334 & 2.5395 & 1.6244 & 0.6835 & 1.8752 & 1.6002 \\
         & p-value                             & $<$0.0001 & $<$0.0001 & 0.2633 & $<$0.0001 & 0.0035 & 0.0597 \\
   \{6\} & $\mbox{RC}_{\nu_{j(i)}}$            & 1.1481 & 0.3146 & 2.0711 & 13.7774 & 0.0845 & 0.2721 \\
         & $\mbox{RC}_{\mbox{SE}(\nu_{j(i)})}$ & 1.0717 & 0.6794 & 0.0710 & 5.8548 & 0.1591 & 0.1516 \\
         & p-value                             & $<$0.0001 & $<$0.0001 & 0.2591 & $<$0.0001 & 0.0025 & 0.0887 \\
   \{9\} & $\mbox{RC}_{\nu_{j(i)}}$            & 6.6762 & 4.8086 & 1.3080 & 1.9930 & 10.5056 & 16.2842 \\
         & $\mbox{RC}_{\mbox{SE}(\nu_{j(i)})}$ & 2.3401 & 4.6059 & 0.7279 & 0.6454 & 1.8089 & 2.3332 \\
         & p-value                             & $<$0.0001 & $<$0.0001 & 0.2461 & $<$0.0001 & 0.0010 & 0.0521 \\
  \{11\} & $\mbox{RC}_{\nu_{j(i)}}$            & 5.9552 & 4.1589 & 0.9594 & 1.3080 & 6.8238 & 15.6473 \\
         & $\mbox{RC}_{\mbox{SE}(\nu_{j(i)})}$ & 2.1207 & 4.0516 & 1.0153 & 0.6422 & 1.8370 & 2.1400 \\
         & p-value                             & $<$0.0001 & $<$0.0001 & 0.2491 & $<$0.0001 & 0.0015 & 0.0529 \\
  \{14\} & $\mbox{RC}_{\nu_{j(i)}}$            & 0.3567 & 0.1849 & 19.1322 & 2.9679 & 4.6735 & 10.1193 \\
         & $\mbox{RC}_{\mbox{SE}(\nu_{j(i)})}$ & 0.2406 & 0.4304 & 3.5654 & 1.1133 & 1.3708 & 1.9766 \\
         & p-value                             & $<$0.0001 & $<$0.0001 & 0.1847 & $<$0.0001 & 0.0018 & 0.1319 \\
  \{28\} & $\mbox{RC}_{\nu_{j(i)}}$            & 3.8931 & 3.0017 & 0.7810 & 1.6211 & 6.2633 & 9.4813 \\
         & $\mbox{RC}_{\mbox{SE}(\nu_{j(i)})}$ & 1.7758 & 3.2991 & 0.5321 & 0.4795 & 1.3797 & 1.6690 \\
         & p-value                             & $<$0.0001 & $<$0.0001 & 0.2551 & $<$0.0001 & 0.0015 & 0.0657 \\
  \{62\} & $\mbox{RC}_{\nu_{j(i)}}$            & 2.4544 & 0.2317 & 17.2524 & 4.0081 & 2.3116 & 9.7044 \\
         & $\mbox{RC}_{\mbox{SE}(\nu_{j(i)})}$ & 0.7470 & 0.4728 & 0.1128 & 0.7484 & 0.3410 & 0.3399 \\
         & p-value                             & $<$0.0001 & $<$0.0001 & 0.3405 & $<$0.0001 & 0.0032 & 0.1240 \\
  \{68\} & $\mbox{RC}_{\nu_{j(i)}}$            & 0.1492 & 0.3021 & 1.9317 & 10.0868 & 0.7457 & 2.3682 \\
         & $\mbox{RC}_{\mbox{SE}(\nu_{j(i)})}$ & 0.5829 & 0.0464 & 0.0464 & 5.8795 & 0.0634 & 0.0850 \\
         & p-value                             & $<$0.0001 & $<$0.0001 & 0.2579 & $<$0.0001 & 0.0026 & 0.0949 \\
  \{92\} & $\mbox{RC}_{\nu_{j(i)}}$            & 5.3858 & 3.4375 & 3.6343 & 2.3131 & 6.1674 & 9.3714 \\
         & $\mbox{RC}_{\mbox{SE}(\nu_{j(i)})}$ & 1.8626 & 3.3714 & 0.6349 & 0.6576 & 1.4118 & 1.7256 \\
         & p-value                             & $<$0.0001 & $<$0.0001 & 0.2695 & $<$0.0001 & 0.0015 & 0.0661 \\
 \{125\} & $\mbox{RC}_{\nu_{j(i)}}$            & 0.8984 & 0.6199 & 33.6173 & 4.5336 & 12.3489 & 11.8086 \\
         & $\mbox{RC}_{\mbox{SE}(\nu_{j(i)})}$ & 0.0543 & 0.0984 & 8.0266 & 1.4191 & 4.6786 & 2.8638 \\
         & p-value                             & $<$0.0001 & $<$0.0001 & 0.1538 & $<$0.0001 & 0.0113 & 0.0632 \\
 \{126\} & $\mbox{RC}_{\nu_{j(i)}}$            & 0.7894 & 0.3005 & 17.9258 & 2.3143 & 5.1040 & 5.9834 \\
         & $\mbox{RC}_{\mbox{SE}(\nu_{j(i)})}$ & 0.2181 & 0.0042 & 0.7860 & 0.4513 & 0.4750 & 0.1782 \\
         & p-value                             & $<$0.0001 & $<$0.0001 & 0.3477 & $<$0.0001 & 0.0015 & 0.1074 \\
 \{155\} & $\mbox{RC}_{\nu_{j(i)}}$            & 0.6267 & 1.1798 & 30.2296 & 0.1697 & 6.2293 & 4.4646 \\
         & $\mbox{RC}_{\mbox{SE}(\nu_{j(i)})}$ & 0.6618 & 1.7310 & 4.5377 & 0.8401 & 2.2139 & 2.5319 \\
         & p-value                             & $<$0.0001 & $<$0.0001 & 0.1508 & $<$0.0001 & 0.0017 & 0.0816 \\
 \{190\} & $\mbox{RC}_{\nu_{j(i)}}$            & 4.7315 & 2.6210 & 8.8509 & 3.5226 & 5.5838 & 2.0483 \\
         & $\mbox{RC}_{\mbox{SE}(\nu_{j(i)})}$ & 1.5743 & 2.6103 & 0.2028 & 0.6890 & 0.9269 & 1.2904 \\
         & p-value                             & $<$0.0001 & $<$0.0001 & 0.2942 & $<$0.0001 & 0.0015 & 0.0850 \\
 \{203\} & $\mbox{RC}_{\nu_{j(i)}}$            & 0.4418 & 0.1479 & 19.1232 & 3.0229 & 4.8032 & 10.1304 \\
         & $\mbox{RC}_{\mbox{SE}(\nu_{j(i)})}$ & 0.2563 & 0.4610 & 3.5651 & 1.1227 & 1.3781 & 1.9903 \\
         & p-value                             & $<$0.0001 & $<$0.0001 & 0.1847 & $<$0.0001 & 0.0018 & 0.1320 \\
 \{205\} & $\mbox{RC}_{\nu_{j(i)}}$            & 0.0180 & 2.1064 & 7.0915 & 0.3799 & 6.2955 & 11.7476 \\
         & $\mbox{RC}_{\mbox{SE}(\nu_{j(i)})}$ & 2.4970 & 0.4491 & 0.2286 & 0.2335 & 0.6286 & 0.5922 \\
         & p-value                             & $<$0.0001 & $<$0.0001 & 0.2851 & $<$0.0001 & 0.0048 & 0.1337 \\
 \{209\} & $\mbox{RC}_{\nu_{j(i)}}$            & 1.8041 & 2.0214 & 9.9710 & 0.8394 & 7.2795 & 13.3072 \\
         & $\mbox{RC}_{\mbox{SE}(\nu_{j(i)})}$ & 3.6545 & 0.6415 & 0.2373 & 0.4213 & 0.4518 & 0.2111 \\
         & p-value                             & $<$0.0001 & $<$0.0001 & 0.3003 & $<$0.0001 & 0.0052 & 0.1392 \\
\{All\} & $\mbox{RC}_{\nu_{j(i)}}$            & 60.1541 & 29.1455 & 42.3398 & 65.9195 & 49.7668 & 34.4222 \\
         & $\mbox{RC}_{\mbox{SE}(\nu_{j(i)})}$ & 36.5916 & 40.6563 & 39.2314 & 29.8468 & 31.7057 & 32.9403 \\
         & p-value                             & $<$0.0001 & $<$0.0001 & 0.2384 & $<$0.0001 & 0.0006 & 0.0839 \\
\hline
\end{tabular} }
\label{influencia_ambiente}
\end{table}

\subsection{Adjustment to the operation characteristics group}

In this last fitting, only the WFP variable is significant in Step 1, the effect of time is measured only by $\alpha_0$ and the GTDL model (without frailty) is the one that best fits these data. Its estimation results are presented in Table \ref{ajuste_operacao}, from which we note that all parameters are significant at the 10\% level.

\begin{table}[H]
\centering
\caption{Estimation results of the GTDL model fitted to the operation characteristics group.}
\resizebox{\linewidth}{!}{
\setlength{\tabcolsep}{3pt}
\begin{tabular}{l|ccc}
  \hline
                 Parameter & MLE     & SE  &  90\% CI \\
  \hline
$\alpha_0$        &  0.1403 & 0.0708 & (0.0238; 0.2568) \\
$\beta_0$         & -2.1940 & 0.5532 & (-3.1040; -1.2840) \\
$\beta_1$ (WFP)  & -0.4236 & 0.1454 & (-0.6627; -0.1845) \\
   \hline
\end{tabular} }
\label{ajuste_operacao}
\end{table}

\begin{figure}[h]
\centering
\includegraphics[width=0.5\linewidth]{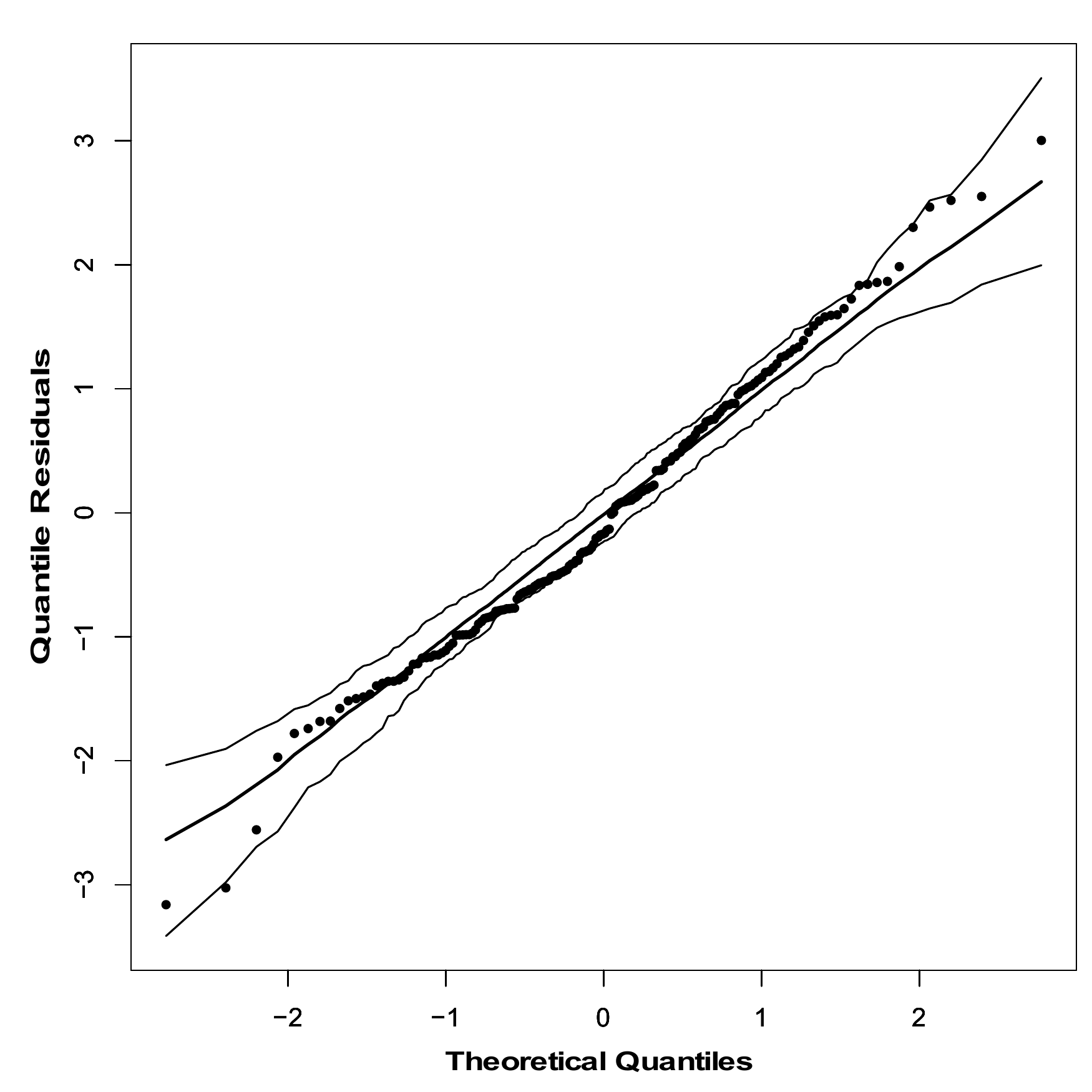}
\caption{QQ-plot with envelope of 95\% for the RQ residuals of the adjustment to the operation characteristics group.}
\label{residuo_operacao}
\end{figure}

Figure \ref{residuo_operacao} shows the QQ-plot of the RQ residuals. In general, we observed a good agreement between the residuals and the standard normal distribution.

\begin{figure}[!]
\centering
\includegraphics[width=1\linewidth]{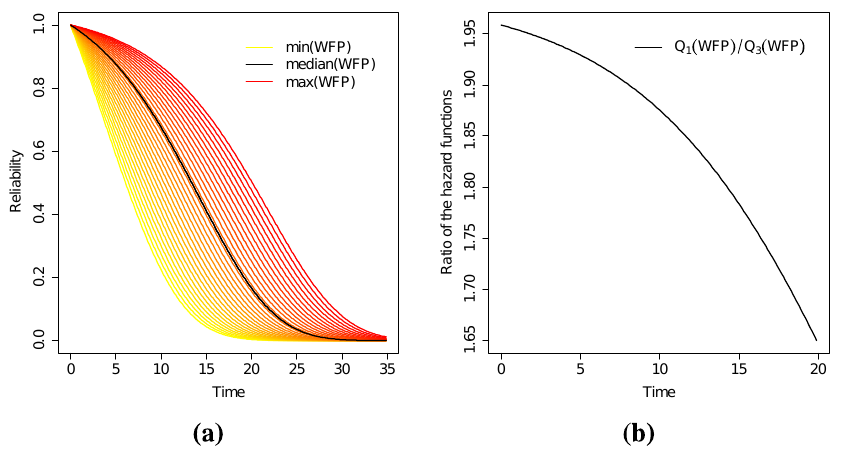}
\caption{(a) Reliability function with variation in the value of the WFP variable. (b) Ratio of hazard functions between the first and third quartiles of the WFP variable.}
\label{conf_operacao}
\end{figure}

Figure \ref{conf_operacao} (a) exhibits the behavior of the reliability function when we vary the value of the WFP variable. Note that with the increase in the value of the WFP variable, an increase in the value of the valve reliability is observed. An example of the HR is shown in Figure \ref{conf_operacao} (b), using the first and third quartiles of the WFP variable, from which we see that the ratio is greater than one, so the risk of failure is greater when the WFP variable takes on the value equal to the first quartile. We also observe that the ratio is initially close to the value 2, indicating that the risk of valve failure, when operated at the value of the first quartile, is approximately twice when operated at the value of the third quartile. Such a risk value decreases with time, but stays above 1.5 in the time of 20 years.

\begin{figure}[!h]
\centering
\includegraphics[width=1\linewidth]{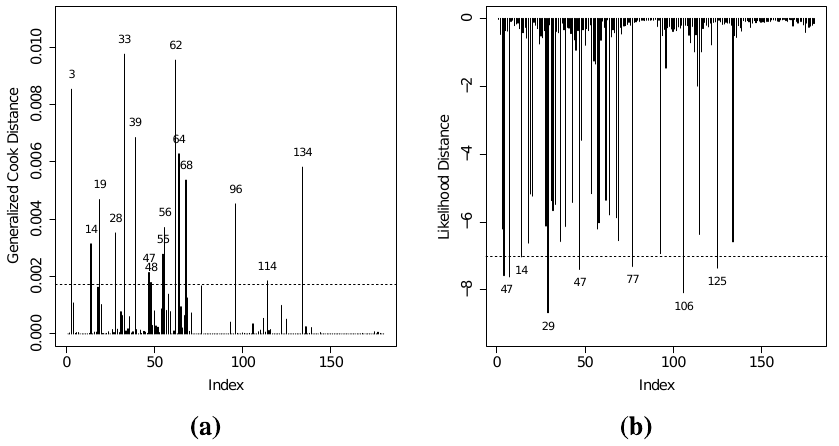}
\caption{(a) Generalized Cook's distance. (b) Likelihood distance, considering the GTDL model fitted to the operation group data.}
\label{cooks_operacao}
\end{figure}

The total number of influential observations detected by the GD and LD measures are 16 and 8, respectively, as can be seen in Figure \ref{cooks_operacao}. Note, however, that only observations 14 and 47 were identified by both metrics. From the removal of influential observations, we can see in Table \ref{influencia_operacao} that the WFP variable remains significant in all configurations, and that the effect of time is almost always significant. The point estimates underwent few changes when excluding only one influential observation; this fact is quantified by RC less than or equal to 20.0448\%. But when we removed all the influential observations, we noticed major changes in the estimates of the parameters $\alpha_0$ and $\beta_1$, with RC of 266\% and 165\%, respectively. In this case, it is worth noting that the estimate of the parameter $\alpha_0 $ changed from 0.1403 to 0.5145, thus the effect of time is greater when removing the influential observations.

\begin{table}[H]
\centering
\caption{The RC values (in \%) for the MLEs and SEs, in addition to the p-values, considering the deleted observations.}
\resizebox{\linewidth}{!}{
\setlength{\tabcolsep}{3pt}
\begin{tabular}{clccccccc}
  \hline
  Deleted case & & $\hat{\alpha}_0$ & $\hat{\beta}_0$ & $\hat{\beta}_1$ & Deleted case & $\hat{\alpha}_0$ & $\hat{\beta}_0$ & $\hat{\beta}_1$\\
  \hline
\{3\}     & $\mbox{RC}_{\nu_{j(i)}}$            &4.5078   & 7.2306 & 8.5400& \{56\}    & 1.4690   & 4.7637 & 6.0048 \\
          & $\mbox{RC}_{\mbox{SE}(\nu_{j(i)})}$ &6.8230   & 0.9056 & 3.8450&           & 0.0069   & 0.9197 & 1.3361 \\
          & p-value                             &0.0765   & 0.0003 & 0.0003&           & 0.0509   & 0.0002 & 0.0023 \\
\{4\}     & $\mbox{RC}_{\nu_{j(i)}}$            &11.0495  & 2.7706 & 2.8200& \{62\}    & 7.0448   & 7.6786 & 8.1094 \\
          & $\mbox{RC}_{\mbox{SE}(\nu_{j(i)})}$ &0.6797   & 1.4186 & 1.8685&           & 0.7200   & 3.6519 & 1.8788 \\
          & p-value                             &0.0289   & 0.0001 & 0.0033&           & 0.0302   &$<$0.0001 & 0.0086 \\
\{7\}     & $\mbox{RC}_{\nu_{j(i)}}$            &4.7590   & 0.3162 & 5.6847& \{64\}    & 8.4898   & 6.2349 & 6.1292 \\
          & $\mbox{RC}_{\mbox{SE}(\nu_{j(i)})}$ &1.0370   & 0.8085 & 1.8114&           & 4.1284   & 1.9374 & 4.1063 \\
          & p-value                             &0.0399   & 0.0001 & 0.0025&           & 0.0816   & 0.0003 & 0.0030 \\
\{14\}    & $\mbox{RC}_{\nu_{j(i)}}$            &8.3125   & 4.2790 & 7.1574& \{68\}    & 6.4199   & 5.7994 & 4.9107 \\
          & $\mbox{RC}_{\mbox{SE}(\nu_{j(i)})}$ &1.8564   & 0.2452 & 1.6713&           & 0.7443   & 2.5868 & 1.4885 \\
          & p-value                             &0.0745   & 0.0002 & 0.0021&           & 0.0363  &$<$0.0001& 0.0063 \\
\{19\}    & $\mbox{RC}_{\nu_{j(i)}}$            &1.6059   & 5.3560 & 6.6352& \{77\}    & 3.4080   & 3.0574 & 7.1069 \\
          & $\mbox{RC}_{\mbox{SE}(\nu_{j(i)})}$ &0.9277   & 3.6788 & 2.0471&           & 1.4973   & 0.3660 & 1.7326 \\
          & p-value                             &0.0461   & 0.0001 & 0.0077&           & 0.0593   & 0.0001 & 0.0022 \\
\{28\}    & $\mbox{RC}_{\nu_{j(i)}}$            &16.0427  & 4.6498 & 3.9293& \{96\}    & 13.9211  & 5.2348 & 5.9877 \\
          & $\mbox{RC}_{\mbox{SE}(\nu_{j(i)})}$ &3.1449   & 0.3587 & 0.9763&           & 4.9171   & 3.4998 & 1.2367 \\
          & p-value                             &0.1068   & 0.0002 & 0.0027&           & 0.0314   & 0.0001 & 0.0068 \\
\{29\}    & $\mbox{RC}_{\nu_{j(i)}}$            &10.6234  & 0.6785 & 8.9663& \{106\}   & 5.8258   & 1.2303 & 7.8611 \\
          & $\mbox{RC}_{\mbox{SE}(\nu_{j(i)})}$ &0.8946   & 0.6947 & 2.1941&           & 1.0602   & 0.6240 & 2.0002 \\
          & p-value                             &0.0298   & 0.0001 & 0.0019&           & 0.0380   & 0.0001 & 0.0021 \\
\{33\}    & $\mbox{RC}_{\nu_{j(i)}}$            &7.9757   & 7.7748 & 7.8140& \{114\}   & 15.4408  & 3.3549 & 2.7208 \\
          & $\mbox{RC}_{\mbox{SE}(\nu_{j(i)})}$ &0.6730   & 3.4475 & 1.7762&           & 4.1372   & 0.9527 & 0.5784 \\
          & p-value                             &0.0336  &$<$0.0001& 0.0083&           & 0.0281  &$<$0.0001& 0.0044 \\
\{39\}    & $\mbox{RC}_{\nu_{j(i)}}$            &9.6157   & 6.5767 & 4.8055& \{125\}   & 0.2446   & 1.6008 & 6.0331 \\
          & $\mbox{RC}_{\mbox{SE}(\nu_{j(i)})}$ &0.6014   & 2.5690 & 1.5782&           & 1.2595   & 0.5882 & 1.7066 \\
          & p-value                             &0.0309  &$<$0.0001& 0.0063&           & 0.0498   & 0.0001 & 0.0024 \\
\{47\}    & $\mbox{RC}_{\nu_{j(i)}}$            &12.0147  & 3.8095 & 1.3540& \{134\}   & 15.7067  & 5.9306 & 6.8143 \\
          & $\mbox{RC}_{\mbox{SE}(\nu_{j(i)})}$ &0.5977   & 1.6467 & 1.8407&           & 2.8567   & 0.3051 & 1.6271 \\
          & p-value                             &0.0274   & 0.0001 & 0.0037&           & 0.1014   & 0.0002 & 0.0022 \\
\{48\}    & $\mbox{RC}_{\nu_{j(i)}}$            &20.0448  & 3.1934 & 8.9491& \{All\} & 266.6522 & 9.7659 & 165.3021 \\
          & $\mbox{RC}_{\mbox{SE}(\nu_{j(i)})}$ &2.9911   & 0.9585 & 0.6384&           & 123.4111 & 50.1084& 108.3383 \\
          & p-value                             &0.0209   & 0.0001 & 0.0016&           & 0.0011   & 0.0171 & 0.0002 \\
\{55\}    & $\mbox{RC}_{\nu_{j(i)}}$            &1.2474   & 4.1259 & 5.0738&           &          &        &         \\
          & $\mbox{RC}_{\mbox{SE}(\nu_{j(i)})}$ &0.0210   & 0.6495 & 0.9201&           &          &        &          \\
          & p-value                             &0.0503   & 0.0002 & 0.0024&           &          &        &         \\
\hline
\end{tabular} }
\label{influencia_operacao}
\end{table}

\section{Conclusions} 
\label{conclusions}

In this paper, we analyzed a real reliability data set on DHSVs used by the Brazil's Petrobras oil firm. This kind of valve has a high reliability, attested by the various technical standards that regulate the oil and gas production sector, consequently few failures are expected during its use. In the graphical analysis, we verified the presence of non-PH. But for some covariates, the Shoenfeld test did not confirm this result.
Then, our proposed modeling was developed using the GTDL and GTDL gamma frailty models with regression also in the parameter that measures the effect of time. 
The decision to use modeling with frailty is in the sense that the variance of the distribution can indicate the absence of important covariates in the modeling, thus indicating that covariates that are in other groups or that for some reason were not registered in the database can be important to explain the time to failure.
The modeling was divided into four groups due to the large amount of missing data. We identified that the variables H2S, BSW, PC, Mfr., Family, OU, CWT, WC and WFP are relevant to describe the time until failure. We also noted that only variables with valve characteristics are not enough to describe the time until failure, because the model with fragility needed to be adopted. 
The residual analysis indicated a good fit of the model to the data, in all groups of covariates.
The global influence analysis highlighted possible influential points in the adjustments made. We presented summaries of the adjustments without the influential observations, because further information and investigations of these observations are for the exclusive use of the company. In this way, we demonstrated the importance of diagnostic analysis in the model because we detected inferential changes after eliminating potentially influential cases.

The variables were divided into 4 groups considering their characteristics and the proposed modeling adopted this division, since the amount of missing data is large. In this work, our interest was to identify possible factors that may influence the time until failure and not to indicate a single model adjusted to a certain group of covariates as being the one that best suits failure times.
As a future work, a study on how to circumvent problems for small sample sizes in the used models using Bayesian methods or bias correction approaches,
the use of the premise that valves installed in the same production regions share the same frailty, thus characterizing the use of shared frailty models, and the use of the well-known statistical method of Principal Component Analysis (PCA) can be modeling alternatives.

\onecolumn

\appendix

\section{}

\begin{figure}[H]
\centering
\includegraphics[width=1\linewidth]{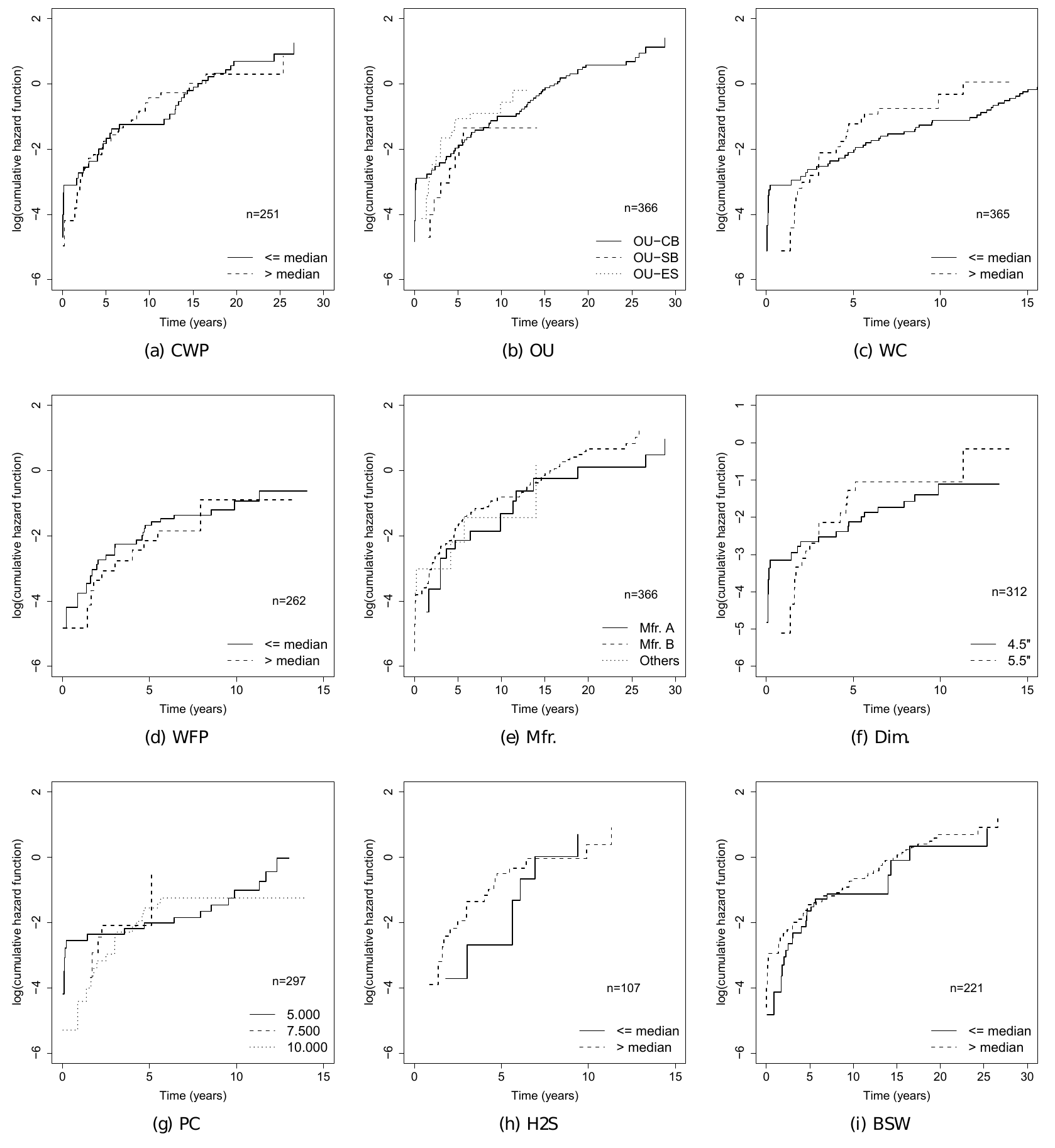}
\vspace*{0.5cm}
\caption{Logarithm of the estimated cumulative hazard function for the variables: (a) CWP, (b) OU, (c) WC, (d) WFP, (e) Mfr., (f) Dim., (g) PC, (h) H2S and (i) BSW. }
\label{verificacao_proporcionalidade_2}
\end{figure}

\twocolumn

\begin{minipage}{0.3\linewidth}

\end{minipage}

\vspace*{-0.6cm}

\begin{wrapfigure}{L}{0.2\textwidth}
\centering
\includegraphics[width=0.2\textwidth, height=5cm]{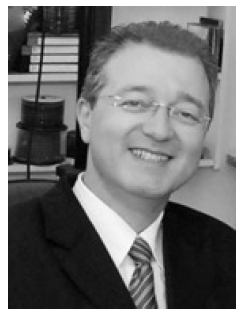}
\end{wrapfigure}
\noindent
Francisco Louzada is a Full Professor of Statistics at the Institute of  Mathematics and Computer Science, University of São Paulo (USP), Brazil. He received his Ph.D. degree in Statistics from the University of Oxford, UK, his M.Sc. degree in Computational Mathematics from USP, Brazil, and his B.Sc. degree in Statistics from UFSCar, Brazil. His main research interests are in survival analysis, data mining, Bayesian inference, classical inference, and probability distribution theory. \\

\vspace*{0.5cm}

\begin{wrapfigure}{L}{0.2\textwidth}
\centering
\includegraphics[width=0.2\textwidth, height=5cm]{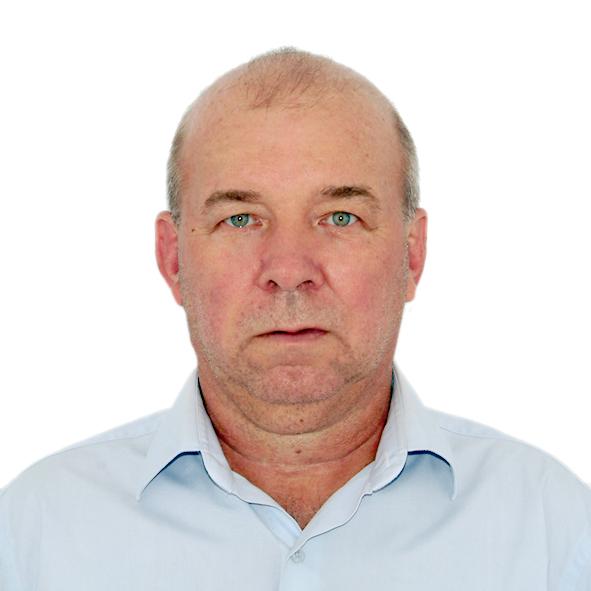}
\end{wrapfigure}
\noindent
Jose Alberto Cuminato is a Full professor in Numerical Analysis at the Institute of Mathematics and Computer Science, University of São Paulo. He received his Ph. D. Degree in Numerical Analysis from the University of Oxford. His main research interests are in Numerical Simulation of Fluid Flows, especially nonNewtonian fluids and also Numerical Linear Algebra. \\

\vspace*{0.5cm}

\begin{wrapfigure}{L}{0.2\textwidth}
\centering
\includegraphics[width=0.2\textwidth, height=5cm]{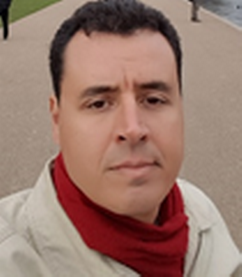}
\end{wrapfigure}
\noindent
Oscar Mauricio Hernandez Rodriguez is Associate Professor of Fluid Mechanics of the São Carlos School of Engineering of the University of São Paulo (USP). MSc and DSc in Mechanical Engineering (Thermo Fluid Sciences, Multiphase Flow), University of Campinas - UNICAMP. Author of more than 150 complete scientific articles in indexed journals or conferences. Head of the Multiphase Flow Research Center of the USP. His main research interests are multiphase flow, hydrodynamic Instability, experimentation and phenomenological modeling of two-phase flow and Instrumentation. \\

\newpage 

\begin{wrapfigure}{L}{0.2\textwidth}
\centering
\includegraphics[width=0.2\textwidth, height=5cm]{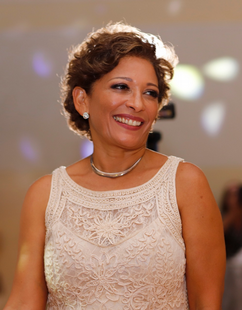}
\end{wrapfigure}
\noindent
Vera Tomazella is Full Professor of Statistics at University Federal of São Carlos, Brazil. Her main interests are survival analysis, reliability analysis, Applied Statistical Genetics, Regression Model Bayesian inference and classical inference. She is Ph.D. from University of São Paulo, Brasil. Postdoctoral (Bayesian Reference Analysis), Universitat de València- Spain and Postdoctoral (Survival Analysis), University of Manchester, Manchester-UK.\\

\vspace*{0.5cm}

\begin{wrapfigure}{L}{0.2\textwidth}
\centering
\includegraphics[width=0.2\textwidth, height=5cm]{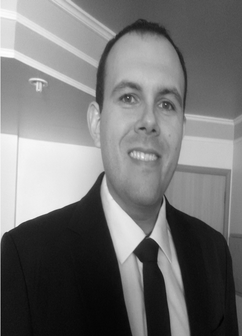}
\end{wrapfigure}
\noindent
Eder Angelo Milani is a Professor of Statistics at the Institute of Mathematics and Statistics, Federal University de Goiás (UFG), Brazil. He received his Ph.D and M.Sc. degree in Statistics from Federal University of São Carlos (UFSCar), Brazil and his B.Sc. degree in Mathematics from São Paulo State University. His main reseach interests are in survival analysis, reliability analysis, classical inference, and time series.\\

\vspace*{0.5cm}

\begin{wrapfigure}{L}{0.2\textwidth}
\centering
\includegraphics[width=0.2\textwidth, height=5cm]{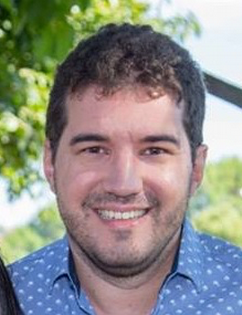}
\end{wrapfigure}
\noindent
Paulo Henrique Ferreira is a Professor of Statistics at the Institute of Mathematics and Statistics, Federal University of Bahia (UFBA), Brazil. He received his Ph.D., M.Sc. and B.Sc. degrees in Statistics all from the Federal University of São Carlos (UFSCar), Brazil. He also has a Postdoctoral training at the University of São Paulo (USP), Brazil. His main research interests include survival and reliability analysis, data mining and statistical process control.\\

\newpage 

\begin{wrapfigure}{L}{0.2\textwidth}
\centering
\includegraphics[width=0.2\textwidth, height=5cm]{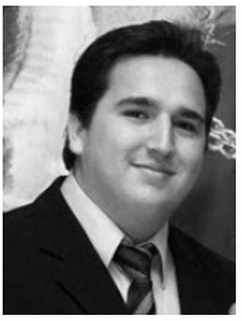}
\end{wrapfigure}
\noindent
Pedro Luiz Ramos is a Postdoctoral Fellow at the Institute of Mathematics and Computer Science, University of São Paulo, Brazil. He received his Ph.D. degree in Statistics from the University of São Paulo, his M.S. degree in Applied and Computational Mathematic in 2014 from the São Paulo State University, Brazil, and his B.S. degree in Statistics in 2011 from the same university. His main research interests are in survival analysis, Bayesian inference, classical inference, and probability distribution theory.\\

\vspace*{0.5cm}

\begin{wrapfigure}{L}{0.2\textwidth}
\centering
\includegraphics[width=0.2\textwidth, height=5cm]{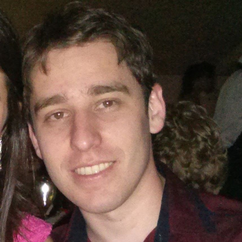}
\end{wrapfigure}
\noindent
Gustavo Bochio is a Postdoctoral Fellow at the Institute of Mathematics and Computer Science, University of São Paulo, Brazil. He received his Ph.D. degree in Mechanical Engineering from the University of São Paulo, his M.S. degree in Naval Engineering in 2015, and his B.S. degree in Mechatronics Engineering in 2011 from the same university. His main research interests include numerical analysis of fluid mechanics, multiphase flows, and turbulence.\\

\vspace*{0.5cm}

\begin{wrapfigure}{L}{0.2\textwidth}
\centering
\includegraphics[width=0.2\textwidth, height=5cm]{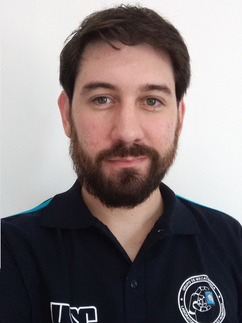}
\end{wrapfigure}
\noindent
Ivan Carlos Perissini is a Mechatronics engineer and Ph.D student in Mechanical Engineering from the University of São Paulo (USP). Received his M.S. degree in Dynamics and mechatronics in 2018 from the same university. His main research interests include computer vision, machine learning and data science. As an engineer worked with product quality for five years at Whirlpool S/A, and is now  working in development projects with Petrobras and USP.\\

\newpage

\begin{wrapfigure}{L}{0.2\textwidth}
\centering
\includegraphics[width=0.2\textwidth, height=5cm]{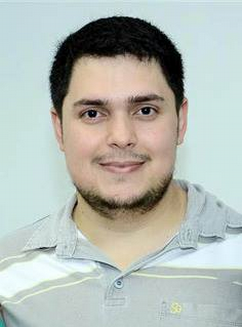}
\end{wrapfigure}
\noindent
Oilson Alberto Gonzatto Junior is a PhD student in Statistics at Federal University of São Carlos (UFSCar) and University of São Paulo (USP), São Carlos, São Paulo, Brazil. He received his M.Sc degree in Biostatistics in 2017 and B.Sc degree in Statistics in 2016 both from State University of Maringá (UEM), Maringá, Paraná, Brazil, his licentiate degree in Mathematics in 2014 from State University of Paraná (UNESPAR). Currently researches in survival and reliability analysis.\\

\vspace*{0.5cm}

\begin{wrapfigure}{L}{0.2\textwidth}
\centering
\includegraphics[width=0.2\textwidth, height=5cm]{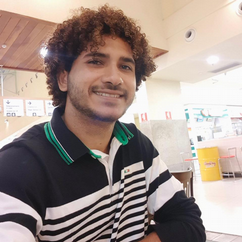}
\end{wrapfigure}
\noindent
Alex Leal Mota is a PhD student in Statistics at Federal University of São Carlos and São Paulo University, São Carlos, Brazil. He received his M.Sc degree in Mathematics in 2017 from the Federal University of Amazonas (UFAM), Manaus, Brazil, and his B.Sc degree in Mathematics in 2014 from Federal University of Amapá (UNIFAP), Macapá, Brazil. His main research interests are in survival and reliability analyses, spatio-temporal models, probability distribution theory, and classical and Bayesian inferences.\\

\vspace*{0.5cm}

\begin{wrapfigure}{L}{0.2\textwidth}
\centering
\includegraphics[width=0.2\textwidth, height=5cm]{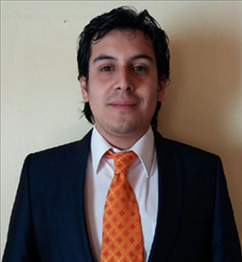}
\end{wrapfigure}
\noindent
Luis Felipe Acuña Alegría is a PhD Student in Engineering Mechanics at São Carlos School of Engineering of the University of São Paulo (USP) in Brazil. He received his MSc degree in Science and Technology of Wood from University of Bio-Bio in Concepción, Chile in 2017. His main research interests are multiphase flow and heat transfer.\\

\newpage

\begin{wrapfigure}{L}{0.2\textwidth}
\centering
\includegraphics[width=0.2\textwidth, height=5cm]{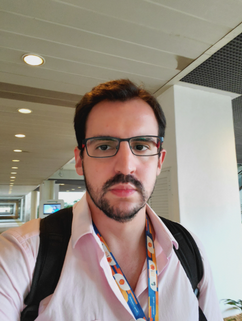}
\end{wrapfigure}
\noindent
Danilo Colombo is a Mechatronics engineer and works as Petroleum engineer at PETROBRAS R\&D Center. He is currently a Ph.D student in Production Engineering from the Universidade Federal Fluminense (UFF). Received his M.S degree in Reliability and Integrity in 2018 from the same university. He has coordinated more than twenty projects in reliability and integrity area for wells engineering applications.\\

\vspace*{0.5cm}

\begin{wrapfigure}{L}{0.2\textwidth}
\centering
\includegraphics[width=0.2\textwidth, height=5cm]{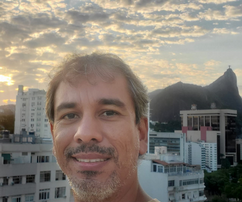}
\end{wrapfigure}
\noindent
Paulo Guilherme Oliveira de Oliveira is a petroleum engineer At Petrobras, Brazil, and currently works as a consultant for oil well valves. He is also a professor at the Veiga de Almeida University (UVA) in Rio de Janeiro, Brazil. Graduated in Mechanical Engineering, he received his M.Sc degree in Mechanical Engineering from PUC-RJ in 2016, and is currently a Dsc student, also from PUC-RJ. At Petrobras, he worked as a company man and currently works in new technology area at the company's headquarter.\\

\vspace*{0.5cm}

\begin{wrapfigure}{L}{0.2\textwidth}
\centering
\includegraphics[width=0.2\textwidth, height=5cm]{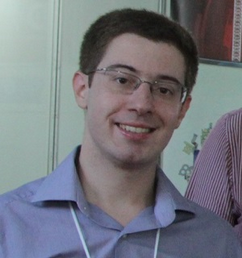}
\end{wrapfigure}
\noindent
Hugo Francisco Lisboa Santos is a Petroleum Engineer at Petrobras, Brazil. He received an M. Sc. degree from the Catholic University of Rio de Janeiro, (PUC-Rio) and a B. Sc. degree from the Military Institute of Engineering (IME). Hugo also has a postgraduate course in Petroleum Engineering at Petrobras. At Petrobras, he works with well engineering and robotics. Hugo won several R\&D prizes, such as the FAA GPS Prize, the 6th Caixa Innovation Prize and the Energy Future Prize.\\

\newpage

\begin{wrapfigure}{L}{0.2\textwidth}
\centering
\includegraphics[width=0.2\textwidth, height=5cm]{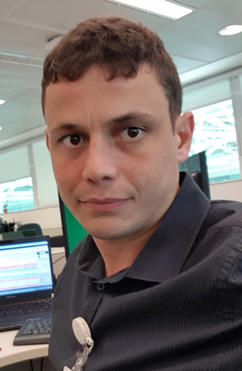}
\end{wrapfigure}
\noindent
Marcus Vinicius de Campos Magalhães is a Petroleum Engineer at Petrobras, Brazil. He completed a postgraduate course in Petroleum Engineering at Petrobras in 2010, and he received his B.Sc. degree in Electrical Engineering in 2006 from Universidade Federal de Goiás (UFG), Brazil. At Petrobras, his experience has spanned from oil and gas production to Company man at offshore rig and now he’s working at Research \& Development Center (CENPES) in Reliability and Integrity.

\EOD

\end{document}